\documentclass[12pt,a4paper]{article}
\usepackage{a4wide}
\usepackage{amsmath}
\usepackage{amssymb}
\usepackage{amsfonts}
\usepackage{epsfig}
\usepackage{subfigure}
\usepackage{exscale}
\usepackage{float}
\usepackage{bbm}
\usepackage[numbers,sort&compress]{natbib}

\newcommand{\R}{{\mathbb{R}}}
\newcommand{\CP}{{\mathbb{C}P}}
\newcommand{\1}{{\mathbbm{1}}}
\newcommand{\p}{\partial}
\newcommand{\eff}{\text{eff}}
\newcommand{\ontopof}[2]{\genfrac{}{}{0pt}{}{#1}{#2}}
\makeatletter
\@addtoreset{equation}{section}
\makeatother

\title{Homogeneous versus Spiral Phases of Hole-doped Antiferromagnets:
A Systematic Effective Field Theory Investigation}

\author{C.\ Br\"ugger$^a$, C.\ P.\ Hofmann$^b$, F.\ K\"ampfer$^a$, 
M.\ Pepe$^c$, \\ and U.-J.~Wiese$^a$
\\ \\
$^a$ Institute for Theoretical Physics, Bern University \\
Sidlerstrasse 5, CH-3012 Bern, Switzerland \\ \\
$^b$ Facultad de Ciencias, Universidad de Colima \\
Bernal D\'iaz del Castillo 340, Colima C.P.\ 28045, Mexico \\ \\
$^c$ Istituto Nazionale di Fisica Nucleare and \\
Dipartimento di Fisica, Universit\`a di Milano-Bicocca \\
3 Piazza della Scienza, 20126 Milano, Italy \\ \\}

\date{September 28, 2006}

\begin{document} 

\maketitle

\vspace{-1cm}

\begin{abstract} \normalsize

Using the low-energy effective field theory for magnons and holes --- the 
condensed matter analog of baryon chiral perturbation theory for pions and 
nucleons in QCD --- we study different phases of doped antiferromagnets. We 
systematically investigate configurations of the staggered magnetization that 
provide a constant background field for doped holes. The most general 
configuration of this type is either constant itself or it represents a spiral 
in the staggered magnetization. Depending on the values of the low-energy 
parameters, a homogeneous phase, a spiral phase, or an inhomogeneous phase is 
energetically favored. The reduction of the staggered magnetization upon doping
is also investigated.

\end{abstract}
 
\section{Introduction}

The precursors of high-temperature superconductors \cite{Bed86} are doped 
antiferromagnets with a spontaneously broken global $SU(2)_s$ spin symmetry and
with 
magnons as the corresponding Goldstone bosons. The effect of antiferromagnetic 
spin fluctuations on the dynamics of doped holes has been investigated in great
detail in the condensed matter literature 
\cite{Bri70,Hir85,And87,Gro87,Tru88,Shr88,Sch88,Cha89,Neu89,Kan89,Wen89,Fis89,Sac89,Sha90,And90,Sin90,Tru90,Kan90,Els90,Dag90,Cha90,Has91,Aue91,Sar91,Ede91,Arr91,Iga92,Fre92,Has93,Kue93,Psa93,Mor93,Sus94,Chu94,Chu95,Zho95,Chu98,Kar98,Man00,Bru00,Sus04,Kot04}. 
Using a variety of numerical and analytic techniques, a wide range of 
interesting phenomena has been investigated in doped antiferromagnets. In 
particular, it was suggested that spiral phases with an inhomogeneous staggered
magnetization may replace the N\'eel phase of the undoped antiferromagnet even 
at arbitrarily small doping 
\cite{Shr88,Kan89,Kan90,Cha90,Aue91,Sar91,Ede91,Arr91,Iga92,Fre92,Psa93,Mor93,Chu95,Chu98,Zho95,Kar98,Man00,Sus04,Kot04}. 
In a spiral phase
the staggered magnetization develops a helix structure, and the N\'eel
ordered antiferromagnet thus turns into a helimagnet. Other inhomogeneities ---
most important stripes --- have also attracted a lot of 
attention \cite{Tra95}. Unfortunately, away from half-filling, the microscopic 
Hubbard and $t$-$J$ models cannot be simulated reliably due to a severe fermion
sign problem. Also analytic calculations are usually not fully systematic but 
suffer from uncontrolled approximations. Consequently, most results for these 
strongly correlated systems remain, at least to some extent, debatable. While
this may seem unavoidable taking into account the complicated nonperturbative 
dynamics of these systems, a systematic effective field theory approach allows 
us to reach some unambiguous conclusions at least for lightly doped systems. 
While some results of this paper have been derived before using less rigorous 
methods, the effective field theory derivation is still very useful, because it
is reliable and particularly transparent. 

Particle physicists are facing the challenges of strongly correlated systems in
studies of the strong
interactions between quarks and gluons. Just like an undoped antiferromagnet,
the QCD vacuum has a spontaneously broken global symmetry --- in that case the 
$SU(2)_L \otimes SU(2)_R$ chiral symmetry --- which gives rise to three 
Goldstone pions --- the analogs of the magnons in an antiferromagnet. The 
QCD analog of the doped holes carrying electric charge are the nucleons
carrying baryon number. Just as simulating the Hubbard model at non-zero doping
is prevented by a fermion sign problem, simulating QCD at non-zero baryon 
density is prevented by a severe complex action problem. For this reason, 
lattice QCD is presently limited to simulating individual particles propagating
in the QCD vacuum. Although simulations of the QCD vacuum do not suffer from 
the complex action problem, they are still very demanding, especially in the
physical regime of small quark masses. Fortunately, a systematic effective 
field theory \cite{Col69,Cal69,Wei79,Gas85} --- chiral perturbation theory --- 
is extremely
successful in describing the low-energy physics in this regime. In chiral 
perturbation theory not quarks and gluons but pions and nucleons are the 
fundamental degrees of freedom. Although the effective theory is not 
renormalizable, it allows a systematic low-energy expansion with only a finite 
number of a priori unknown low-energy parameters entering at each order. The 
values of the low-energy parameters can be determined from experiments or from
lattice QCD simulations. Chiral perturbation theory provides us with precise 
predictions for the low-energy pion physics, which would be practically 
impossible to derive directly from QCD. Baryon chiral perturbation theory 
\cite{Geo84,Gas88,Jen91,Ber92,Bec99} extends these successes to the low-energy 
physics of both pions and nucleons. At present, a fully systematic 
power-counting scheme seems to exist only for the sector with a single nucleon 
\cite{Bec99}. Still, few-nucleon systems have also been treated quantitatively 
\cite{Wei90,Kap98,Epe98,Bed98,Kol99}. The QCD analog of a spiral phase in a
doped antiferromagnet is a pion condensate in nuclear matter 
\cite{Bah65,Mig71,Saw72,Sca72,Bay73,Cam75}.

The systematic technique of chiral perturbation theory is not limited to QCD 
but can be applied to any system with Goldstone bosons. 
Indeed, systematic low-energy effective theories have been very successful in 
describing the dynamics of magnons in both ferro- and antiferromagnets
\cite{Cha89,Neu89,Fis89,Has90,Has91,Has93,Chu94,Leu94,Hof99,Rom99,Bae04}.
In \cite{Kae05,Bru06} we have extended the pure magnon effective field theory 
by including charge carriers. The resulting effective theory for magnons and 
doped holes is the condensed matter analog of baryon chiral perturbation 
theory. The effective theory incorporates important experimental as well as
theoretical results, such as the location of hole pockets at lattice momenta 
$(\frac{\pi}{2a},\pm \frac{\pi}{2a})$ which follows from angle resolved 
photoemission spectroscopy (ARPES) experiments \cite{Wel95,LaR97,Kim98,Ron98} 
as well as from theoretical investigations of Hubbard or $t$-$J$-like models
\cite{Tru88,Shr88,Els90,Bru00}.
Recently, we have used the effective theory to derive the magnon-mediated 
forces between two isolated holes in an otherwise undoped system 
\cite{Bru05,Bru06}. Remarkably, the 
Schr\"odinger equation corresponding to the one-magnon exchange potential can 
be solved analytically and gives rise to an infinite number of bound states. It
remains to be seen if these isolated hole pairs are related to the preformed 
Cooper pairs of high-temperature superconductivity. In this paper, we use the
effective theory to investigate the regime of small doping. This is possible
analytically if the 4-fermion contact interactions between doped holes
are weak and can be treated perturbatively. Whether this is indeed the case
depends on the concrete magnetic material under consideration. It should be
noted that the 4-fermion couplings between doped holes in the effective theory
may well be small, although the microscopic on-site Coulomb repulsion $U$
in the Hubbard model or the exchange coupling $J$ in the $t$-$J$ model which 
cause antiferromagnetism are strong. In particular, in the effective theory 
antiferromagnetism arises independent of the strength of the remnant 4-fermion
couplings between doped holes. Assuming that the 4-fermion couplings can be
treated perturbatively, the effective theory predicts both homogeneous and 
spiral phases, depending on the specific values of the low-energy parameters.

The paper is organized as follows. In section 2 the effective theory of magnons
and holes as well as the nonlinear realization of the spontaneously broken 
$SU(2)_s$ spin symmetry are reviewed. The holes interact with the Goldstone 
bosons via a $U(1)_s$ ``gauge'' field and two ``charged'' vector fields 
composed of
magnons. The gauge group $U(1)_s$ and the corresponding ``charge'' refer to the
unbroken subgroup of $SU(2)_s$. In section 3 we consider the most general 
magnon field that gives rise to constant gauge and charged vector fields and 
thus to a homogeneous background for the doped holes. These magnon fields turn 
out either to be homogeneous themselves or to form a spiral in the staggered
magnetization. A particular magnon field which gives rise to inhomogeneous
composite gauge and charged vector fields --- a so-called double spiral --- is
also discussed. In section 4 homogeneous and spiral phases are investigated. 
The effect of weak 4-fermion contact interactions is investigated in section 5 
using perturbation theory and --- depending on the values of the low-energy 
parameters --- it is determined which phase is energetically favored. In
section 6 the reduction of the staggered magnetization upon doping is 
calculated both for the homogeneous and spiral phases. Section 7 contains an 
outlook as well as our conclusions. In an appendix we prove that the most
general configuration of the staggered magnetization that provides a constant
background field for the doped holes is either constant itself or represents a
spiral.

\section{Systematic Low-Energy Effective Field Theory for Magnons and Holes}

In order to make this paper self-contained, in this section we review the
effective theory for magnons and holes constructed in \cite{Kae05,Bru06} which 
is based on the pure magnon effective theory of
\cite{Neu89,Fis89,Has90,Has91,Has93,Leu94,Hof99,Chu94,Rom99}.

The staggered magnetization of an antiferromagnet is described by a unit-vector
field
\begin{equation}
\vec e(x) =
(\sin\theta(x) \cos\varphi(x),\sin\theta(x) \sin\varphi(x),\cos\theta(x)),
\end{equation}
in the coset space $SU(2)_s/U(1)_s = S^2$, with $x = (x_1,x_2,t)$ 
denoting a point in $(2+1)$-dimensional space-time. For our purposes it is 
more convenient (but completely equivalent) to use a $\CP(1)$ representation in
terms of $2 \times 2$ Hermitean projection matrices $P(x)$ that obey
\begin{equation}
P(x) = \frac{1}{2}(\1 + \vec e(x) \cdot \vec \sigma), \quad
P(x)^\dagger = P(x), \quad \mbox{Tr} P(x) = 1, \quad P(x)^2 = P(x),
\end{equation}
where $\vec \sigma$ are the Pauli matrices. The relevant symmetries are 
realized as follows
\begin{alignat}{3}
SU(2)_s:&\quad &P(x)' &= g P(x) g^\dagger, \nonumber \\
D_i:&\quad &^{D_i}P(x) &= \1 - P(x), \nonumber \\
O:&\quad &^OP(x) &= P(Ox), &\quad Ox &= (- x_2,x_1,t), \nonumber \\
R:&\quad &^RP(x) &= P(Rx), &\quad Rx &= (x_1,- x_2,t), \nonumber \\
T:&\quad &^TP(x) &= \1 - P(Tx), &\quad Tx &= (x_1,x_2,- t).
\end{alignat}
Here $g \in SU(2)_s$ is a matrix that implements the global spin symmetry which
is spontaneously broken down to $U(1)_s$, $D_i$ denotes the displacement by one
lattice spacing in the $i$-direction, and $O$, $R$, and $T$ denote 90 degrees 
spatial rotations, spatial reflections, and time-reversal, respectively.

In order to couple doped holes to the magnons, a nonlinear realization of the 
$SU(2)_s$ symmetry was constructed in \cite{Kae05}. The magnon field is
diagonalized by a unitary transformation $u(x) \in SU(2)_s$, i.e.
\begin{equation}
u(x) P(x) u(x)^\dagger = \frac{1}{2}(\1 + \sigma_3) = 
\left(\begin{array}{cc} 1 & 0 \\ 0 & 0 \end{array} \right), \quad 
u_{11}(x) \geq 0,
\end{equation}
with
\begin{equation}
u(x) = \left(\begin{array}{cc} \cos(\frac{1}{2}\theta(x)) & 
\sin(\frac{1}{2}\theta(x)) \exp(- i \varphi(x)) \\
- \sin(\frac{1}{2}\theta(x)) \exp(i \varphi(x)) & \cos(\frac{1}{2}\theta(x)) 
\end{array}\right).
\end{equation}
The transformation $u(x)$ describes a rotation of the local staggered
magnetization vector $\vec e(x)$ into the 3-direction. Since $u(x)$ is more
directly related to $P(x)$ than to $\vec e(x)$ itself, we have chosen the 
$\CP(1)$ representation. Under a global $SU(2)_s$ transformation $g$ the 
diagonalizing field $u(x)$ transforms as
\begin{equation}
\label{trafou}
u(x)' = h(x) u(x) g^\dagger, \quad u_{11}(x)' \geq 0,
\end{equation}
which defines the nonlinear $U(1)_s$ symmetry transformation 
\begin{equation}
h(x) = \exp(i \alpha(x) \sigma_3),
\end{equation}
that acts like a gauge transformation in the unbroken subgroup $U(1)_s$. The
local symmetry transformation $h(x)$ depends on the global transformation $g$ 
as well as on the local staggered magnetization $P(x)$ from which it inherits
its $x$-dependence. Under the displacement symmetry 
$D_i$ one obtains 
\begin{equation}
\label{taueq}
^{D_i}u(x) = \tau(x) u(x), \quad
\tau(x) = \left(\begin{array}{cc} 0 & - \exp(- i \varphi(x)) \\
\exp(i \varphi(x)) & 0 \end{array} \right).
\end{equation}

The way in which the global $SU(2)_s$ spin symmetry disguises itself as a local
symmetry in the unbroken $U(1)_s$ subgroup is characteristic for any systematic
effective field theory of Goldstone bosons. The nonlinear realization of 
spontaneously broken continuous global symmetries has been discussed in full 
generality in the pioneering work of Callan, Coleman, Wess, and Zumino 
\cite{Col69,Cal69}. Following their general scheme, 
doped holes are derivatively coupled to the magnons. In fact, the holes are
``charged'' under the local $U(1)_s$ symmetry and transform with the nonlinear
transformation $h(x)$. In order to couple holes to the magnons it is necessary
to introduce the anti-Hermitean traceless field
\begin{equation}
v_\mu(x) = u(x) \p_\mu u(x)^\dagger,
\end{equation}
which obeys the following transformation rules
\begin{alignat}{3}
SU(2)_s:&\quad &v_\mu(x)' &= h(x) [v_\mu(x) + \p_\mu] h(x)^\dagger,
  \hspace{-5em} \nonumber \\
D_i:&\quad &^{D_i}v_\mu(x) &= \tau(x)[v_\mu(x) + \p_\mu] \tau(x)^\dagger,
  \hspace{-5em} \nonumber \\
O:&\quad &^Ov_i(x) &= \varepsilon_{ij} v_j(Ox), \quad
  &^Ov_t(x) &= v_t(Ox), \nonumber \\
R:&\quad &^Rv_1(x) &= v_1(Rx), \quad &^Rv_2(x) &= - v_2(Rx),
  \quad ^Rv_t(x) = v_t(Rx), \nonumber \\
T:&\quad &^Tv_j(x) &= \ ^{D_i}v_j(Tx), \quad &^Tv_t(x) &= - \ ^{D_i}v_t(Tx).
\end{alignat}
Writing
\begin{equation}
v_\mu(x) = i v_\mu^a(x) \sigma_a, \quad 
v_\mu^\pm(x) = v_\mu^1(x) \mp i v_\mu^2(x),
\end{equation}
the field $v_\mu(x)$ decomposes into an Abelian ``gauge'' field $v_\mu^3(x)$ 
and two ``charged'' vector fields $v_\mu^\pm(x)$, which transform as
\begin{equation}
v_\mu^3(x)' = v_\mu^3(x) - \p_\mu \alpha(x), \quad 
v_\mu^\pm(x)' = v_\mu^\pm(x) \exp(\pm 2 i \alpha(x)),
\end{equation}
under $SU(2)_s$.

ARPES measurements \cite{Wel95,LaR97,Kim98,Ron98} as well as theoretical
calculations in $t$-$J$-like models \cite{Tru88,Els90,Bru00} have revealed that
at small doping holes occur in pockets centered at 
$k^\alpha = (\frac{\pi}{2a},\frac{\pi}{2a})$ and 
$k^\beta = (\frac{\pi}{2a},- \frac{\pi}{2a})$ in the Brillouin zone. 
The elliptically shaped hole pockets are illustrated in figure 1.
\begin{figure}[t]
\begin{center}
\vspace{-0.4cm}
\epsfig{file=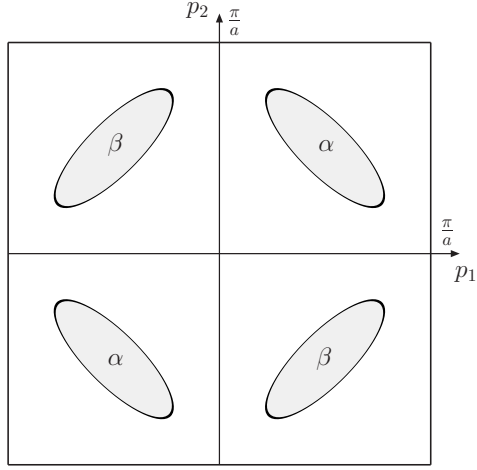,width=7cm}
\end{center}
\caption{\it Elliptically shaped hole pockets centered at 
$(\pm \frac{\pi}{2a},\pm \frac{\pi}{2a})$. Two pockets centered at $k^f$
and $k^f + (\frac{\pi}{a},\frac{\pi}{a})$ combine to form the pockets for the 
flavors $f = \alpha,\beta$.}
\end{figure}
The effective field theory is defined in the space-time continuum and the holes
are described by Grassmann-valued fields $\psi^f_s(x)$ carrying a ``flavor'' 
index $f = \alpha, \beta$ that characterizes the corresponding hole pocket. The
index $s = \pm$ denotes spin parallel ($+$) or antiparallel ($-$) to the local 
staggered magnetization. Following \cite{Kae05,Bru06}, under the various 
symmetry operations the hole fields transform as
\begin{alignat}{2}
SU(2)_s:&\quad &\psi^f_\pm(x)' &= \exp(\pm i \alpha(x)) \psi^f_\pm(x),
\nonumber \\
U(1)_Q:&\quad &^Q\psi^f_\pm(x) &= \exp(i \omega) \psi^f_\pm(x),
\nonumber \\
D_i:&\quad &^{D_i}\psi^f_\pm(x) &= 
\mp \exp(i k^f_i a) \exp(\mp i \varphi(x)) \psi^f_\mp(x),
\nonumber \\
O:&\quad &^O\psi^\alpha_\pm(x) &= \mp \psi^\beta_\pm(Ox), \quad
^O\psi^\beta_\pm(x) = \psi^\alpha_\pm(Ox), \nonumber \\
R:&\quad &^R\psi^\alpha_\pm(x) &= \psi^\beta_\pm(Rx), \quad\;\;\;
^R\psi^\beta_\pm(x) = \psi^\alpha_\pm(Rx), \nonumber \\
T:&\quad &^T\psi^f_\pm(x) &= \mp \exp(\mp i \varphi(Tx)) 
\psi^{f\dagger}_\pm(Tx),
\nonumber \\
&\quad &^T\psi^{f\dagger}_\pm(x) &= \pm \exp(\pm i \varphi(Tx))
\psi^f_\pm(Tx).
\end{alignat}
Here $U(1)_Q$ is the fermion number symmetry of the holes. Interestingly, in 
the effective continuum theory the location of holes in lattice momentum space 
manifests itself as a ``charge'' $k^f_i$ under the displacement symmetry $D_i$.

Once the relevant low-energy degrees of freedom have been identified, and the
transformation rules of the corresponding fields have been understood, the
construction of the effective action is uniquely determined. The low-energy 
effective action of magnons and holes is constructed as a derivative expansion.
At low energies terms with a small number of derivatives dominate the dynamics.
Since the holes are heavy nonrelativistic fermions, one time-derivative counts 
like two spatial derivatives. Here we limit ourselves to terms with at most one
temporal or two spatial derivatives. One then constructs all terms consistent 
with the symmetries listed above. The effective action 
can be written as
\begin{equation}
S[\psi^{f\dagger}_\pm,\psi^f_\pm,P] = \int d^2x \ dt \ \sum_{n_\psi}
{\cal L}_{n_\psi},
\end{equation}
where $n_\psi$ denotes the number of fermion fields that the various terms 
contain. The leading terms in the pure magnon sector take the form
\begin{eqnarray}
{\cal L}_0&=&\rho_s \mbox{Tr}
\big[ \p_i P \p_i P + \frac{1}{c^2} \p_t P \p_t P \big] \nonumber \\
&=&\frac{\rho_s}{2} \left(\p_i \vec e \cdot \p_i \vec e  + 
\frac{1}{c^2} \p_t \vec e \cdot \p_t \vec e \right) =
2 \rho_s \left(v_i^+ v_i^- + \frac{1}{c^2} v_t^+ v_t^-\right).
\end{eqnarray}
Here $\rho_s$ is the spin stiffness and $c$ is the spinwave velocity. The 
leading terms with two fermion fields (containing at most one temporal or two 
spatial derivatives) are given by
\begin{align}
\label{Lagrangian2}
{\cal L}_2=\sum_{\ontopof{f=\alpha,\beta}{\, s = +,-}} \Big[ &
M \psi^{f\dagger}_s \psi^f_s + \psi^{f\dagger}_s D_t \psi^f_s 
\nonumber \\[-3ex]
&+\frac{1}{2 M'} D_i \psi^{f\dagger}_s D_i \psi^f_s +
\sigma_f \frac{1}{2 M''} \big( D_1 \psi^{f\dagger}_s D_2 \psi^f_s +
D_2 \psi^{f\dagger}_s D_1 \psi^f_s \big) \nonumber \\[0.7ex]
&+\Lambda \big( \psi^{f\dagger}_s v^s_1 \psi^f_{-s} 
+ \sigma_f \psi^{f\dagger}_s v^s_2 \psi^f_{-s} \big) \nonumber \\
&+N_1 \psi^{f\dagger}_s v^s_i v^{-s}_i \psi^f_s +
\sigma_f N_2 \big( \psi^{f\dagger}_s v^s_1 v^{-s}_2 \psi^f_s + 
\psi^{f\dagger}_s v^s_2 v^{-s}_1 \psi^f_s \big) \Big].
\end{align}
It should be noted that $v_i^\pm(x)$ contains a spatial derivative, such that
magnons and holes are indeed derivatively coupled. In eq.({\ref{Lagrangian2}) 
$M$ is the rest mass and $M'$ and $M''$ are the kinetic masses of a hole, 
$\Lambda$ is a hole-one-magnon, and $N_1$ and $N_2$ are hole-two-magnon 
couplings, which all take real values. The sign $\sigma_f$ is $+$ for 
$f = \alpha$ and $-$ for $f = \beta$. The covariant derivative takes the form
\begin{equation}
D_\mu \psi^f_\pm(x) = \p_\mu \psi^f_\pm(x) \pm i v^3_\mu(x) \psi^f_\pm(x).
\end{equation}
The leading terms with four fermion fields are given by
\begin{align}
\label{Lagrange4}
{\cal L}_4 = \sum_{s = +,-} \Big\{ &
\frac{G_1}{2} (\psi^{\alpha\dagger}_s \psi^\alpha_s 
\psi^{\alpha\dagger}_{-s} \psi^\alpha_{-s} + 
\psi^{\beta\dagger}_s \psi^\beta_s 
\psi^{\beta\dagger}_{-s} \psi^\beta_{-s}) \nonumber \\[-1ex]
&+G_2 \psi^{\alpha\dagger}_s \psi^\alpha_s \psi^{\beta\dagger}_s \psi^\beta_s +
G_3 \psi^{\alpha\dagger}_s \psi^\alpha_s 
\psi^{\beta\dagger}_{-s} \psi^\beta_{-s} \nonumber \\[.7ex]
&+G_4 \Big[ \psi^{\alpha\dagger}_s \psi^\alpha_s
\sum_{s' = +,-} \big( \psi^{\beta\dagger}_{s'} v^{s'}_1 \psi^\beta_{-s'} - 
\psi^{\beta\dagger}_{s'} v^{s'}_2 \psi^\beta_{-s'} \big) \nonumber \\
&\hspace{2.7em}+\psi^{\beta\dagger}_s \psi^\beta_s
\sum_{s' = +,-} \big( \psi^{\alpha\dagger}_{s'} v^{s'}_1 \psi^\alpha_{-s'} + 
\psi^{\alpha\dagger}_{s'} v^{s'}_2 \psi^\alpha_{-s'} \big) \Big] \Big\},
\end{align}
with the real-valued 4-fermion coupling constants $G_1$, $G_2$, $G_3$, and 
$G_4$. Here we have limited ourselves to terms containing at most one spatial
derivative. In principle there are even more contact interactions
among the fermions such as 6- and 8-fermion couplings as well as 4-fermion
couplings including more derivatives. Some of these terms have been constructed
in \cite{Bru06} but play no role in the present work and have hence been
suppressed.

Remarkably, the above Lagrangian has an accidental global $U(1)_F$ flavor 
symmetry that acts as
\begin{equation}
U(1)_F: \ ^F\psi^f_\pm(x) = \exp(\sigma_f i \eta) \psi^f_\pm(x).
\end{equation}
This symmetry is not present in the underlying microscopic systems and is
indeed explicitly broken by higher-order terms in the effective action.
For $c \rightarrow \infty$ the leading terms of the effective action also have 
an accidental Galilean boost symmetry
\begin{alignat}{2}
\label{boost}
G:&\quad &^GP(x) &= P(Gx), \qquad Gx = (\vec x - \vec v \ t,t), \nonumber \\
  &\quad &^G \psi^f_\pm(x) &= 
  \exp\left(i \vec p^f \cdot \vec x - \omega^f t\right) \psi^f_\pm(Gx),
\nonumber \\
  &\quad &^G \psi^{f\dagger}_\pm(x) &= 
  \psi^{f\dagger}_\pm(Gx) 
\exp\left(- i \vec p^f \cdot \vec x + \omega^f t\right),
\end{alignat}
with $\vec p^f = (p^f_1, p^f_2)$ and $\omega^f$ given by
\begin{align}
p^f_1 &= \frac{M'}{1 - (M'/M'')^2}
\left(v_1 - \sigma_f \frac{M'}{M''} v_2 \right), \quad
p^f_2 = \frac{M'}{1 - (M'/M'')^2}
\left(v_2 - \sigma_f \frac{M'}{M''} v_1 \right), \nonumber \\
\omega^f &= \frac{{p^f_i}^2}{2 M'} + \sigma_f \frac{p^f_1 p^f_2}{M''} =
\frac{M'}{1 - (M'/M'')^2}\left[\frac{1}{2}(v_1^2 + v_2^2) 
- \sigma_f \frac{M'}{M''} v_1 v_2\right].
\end{align}
Also the Galilean boost symmetry is explicitly broken at higher orders of the 
derivative expansion. In the real materials Galilean (or actually Poincar\'e) 
invariance is spontaneously broken by the formation of a crystal lattice, with 
phonons as the corresponding Goldstone bosons. Here we assume that phonons play
no major role in the cuprates, and we focus entirely on the magnons. Still, 
phonons and a spontaneously broken Galilean symmetry could be included in the 
effective field theory if necessary.

\section{Spirals in the Staggered Magnetization}

In the following we will consider configurations $\vec e(x)$ of the staggered
magnetization which --- although not necessarily constant themselves --- 
provide a constant background field for the doped holes. We restrict ourselves 
to time-independent configurations, such that $v_t(x) = 0$. The most general 
configuration, with $v_i(x)$ constant up to a gauge transformation, represents 
a spiral in the staggered magnetization. We also discuss a so-called double 
spiral which gives rise to a non-uniform composite vector field and thus to an 
inhomogeneous fermion density.

\subsection{Spirals with Uniform Composite Vector Fields}

Since the holes couple to the composite vector field $v_i(x)$ in a gauge 
covariant way, it is sufficient to assume that $v_i(x)$ is constant only up to 
a gauge transformation, i.e.
\begin{eqnarray}
\label{const}
{v^3_i}(x)'&=&v^3_i(x) - \p_i \alpha(x) = 
\sin^2\frac{\theta(x)}{2} \p_i \varphi(x) - \p_i \alpha(x) = c^3_i,
\nonumber \\
{v^\pm_i}(x)'&=&v^\pm_i(x) \exp(\pm 2 i \alpha(x)) \nonumber \\
&=&\frac{1}{2} \big[\sin\theta(x) \p_i \varphi(x) \pm i \p_i \theta(x)\big]
\exp(\mp i (\varphi(x) - 2 \alpha(x))) = c^\pm_i,
\end{eqnarray}
with $c^3_i$ and $c^\pm_i$ being constant. As shown in the appendix, the most 
general configuration that leads to a constant $v_i(x)'$ represents a spiral in
the staggered magnetization. Furthermore, by an appropriate gauge 
transformation one can always achieve
\begin{equation}
c_i^+ = c_i^- = c_i \in \R.
\end{equation}
The magnon contribution to the energy density of these configurations is given 
by
\begin{equation}
\epsilon_m = \frac{\rho_s}{2} \p_i \vec e(x) \cdot \p_i \vec e(x) = 
2 \rho_s v_i^+(x) v_i^-(x) = 2 \rho_s (c_1^2 + c_2^2).
\end{equation}

To be specific, let us consider a concrete family of spiral configurations with
\begin{equation}
\label{special}
\theta(x) = \theta_0, \quad \varphi(x) = k_i x_i,
\end{equation}
which implies
\begin{equation}
v_t(x) = 0, \quad  v^3_i(x) = k_i \sin^2\frac{\theta_0}{2}, \quad
v_i^\pm(x) = \frac{k_i}{2} \sin\theta_0 \exp(\mp i k_i x_i).
\end{equation}
Choosing the gauge transformation
\begin{equation}
\alpha(x) = \frac{1}{2} k_i x_i,
\end{equation}
one obtains the constant field
\begin{eqnarray}
\label{constant}
&&v_t(x)' = 0, \quad 
{v^3_i}(x)' = v^3_i(x) - \p_i \alpha(x) = 
k_i (\sin^2\frac{\theta_0}{2} - \frac{1}{2}) = c^3_i, \nonumber \\
&&{v^\pm_i}(x)' = v^\pm_i(x) \exp(\pm 2 i \alpha(x)) = 
\frac{k_i}{2} \sin\theta_0 = c_i.
\end{eqnarray}
Hence, comparing with the appendix, in this case we can identify
\begin{equation}
\label{ci3}
c^3_i = - \frac{k_i}{2} \cos\theta_0, \quad
a = \frac{c_i}{c^3_i} = - \tan\theta_0,
\end{equation}
and the magnon contribution to the energy density is
\begin{equation}
\epsilon_m = 2 \rho_s (c_1^2 + c_2^2) =
\frac{\rho_s}{2} (k_1^2 + k_2^2) \sin^2\theta_0.
\end{equation}

\subsection{A Double Spiral}

For most of this paper we restrict ourselves to configurations of the staggered
magnetization which give rise to a homogeneous composite vector field 
$v_i(x)'$. However, in this subsection we examine a configuration with 
an inhomogeneous $v_i(x)'$ --- the so-called double spiral \cite{Kan90}. 
Although we will not explore this configuration any further in this work, it is
interesting for future investigations. In particular, one can study it in order
to figure out whether spirals with constant $v_i(x)'$ may be unstable against 
developing inhomogeneities.

A particularly simple form of a double spiral is given by
\begin{equation}
\vec e(x) = 
(\sin(k_1 x_1) \cos(k_2 x_2),\sin(k_2 x_2),\cos(k_1 x_1) \cos(k_2 x_2)).
\end{equation}
The magnon contribution to the energy density of the double spiral takes the
form
\begin{equation}
\epsilon_m = \frac{\rho_s}{2} \p_i \vec e(x) \cdot \p_i \vec e(x) =
\frac{\rho_s}{2} \left(k_1^2 \cos^2(k_2 x_2) + k_2^2\right).
\end{equation}
It is straightforward to compute the composite vector field for the double
spiral and one obtains
\begin{eqnarray}
&&v_1^3(x) = - \frac{k_1 \cos(k_1 x_1) \sin(k_2 x_2) \cos(k_2 x_2)}
{2 (1 + \cos(k_1 x_1) \cos(k_2 x_2))}, \nonumber \\
&&v_1^\pm(x) = \frac{k_1 \cos(k_2 x_2)
[\sin(k_1 x_1) \sin(k_2 x_2) \pm i (\cos(k_1 x_1) + \cos(k_2 x_2))]}
{2 (1 + \cos(k_1 x_1) \cos(k_2 x_2))}, \nonumber \\
&&v_2^3(x) = \frac{k_2 \sin(k_1 x_1)}{2 (1 + \cos(k_1 x_1) \cos(k_2 x_2))}, 
\nonumber \\
&&v_2^\pm(x) = 
\frac{k_2 [\cos(k_1 x_1) + \cos(k_2 x_2) \mp i \sin(k_1 x_1) \sin(k_2 x_2)]}
{2 (1 + \cos(k_1 x_1) \cos(k_2 x_2))}.
\end{eqnarray}
As before we perform a gauge transformation
\begin{equation}
{v^3_i}(x)' = v^3_i(x) - \p_i \alpha(x), \quad
{v^\pm_i}(x)' = v^\pm_i(x) \exp(\pm 2 i \alpha(x)),
\end{equation}
in this case with
\begin{equation}
\exp(2 i \alpha(x)) = 
\frac{\sin(k_1 x_1) \sin(k_2 x_2) - i (\cos(k_1 x_1) + \cos(k_2 x_2))}
{1 + \cos(k_1 x_1) \cos(k_2 x_2)},
\end{equation}
and we obtain the remarkably simple form
\begin{alignat}{2}
{v_1^3}(x)'& = - \frac{k_1}{2} \sin(k_2 x_2),
&\quad {v_1^\pm}(x)' &=  \frac{k_1}{2} \cos(k_2 x_2), \nonumber \\
{v_2^3}(x)' &= 0,
&\quad {v_2^\pm}(x)' &= \mp i \frac{k_2}{2}.
\end{alignat}
\section{Homogeneous versus Spiral Phases}

In this section we calculate the fermionic contribution to the energy density 
of a configuration with constant $v_i(x)'$ in order to decide which 
configuration is energetically favored. It will turn out that this depends on 
the values of the low-energy parameters. For large $\rho_s$ the magnon 
contribution to the energy density dominates, and a homogeneous phase is 
favored. In that case, all four hole pockets are equally populated with doped 
holes of both spin up and spin down. For smaller $\rho_s$, on the other hand, 
the fermionic contribution to the energy density dominates and favors a spiral 
configuration. In this case, only a particular linear combination of spin up 
and spin down states is occupied by doped holes. If the spiral is oriented 
along a crystal lattice axis (a zero degree spiral) hole pockets of both types 
($\alpha$ and $\beta$) are filled with doped holes. On the other hand, if the 
spiral is oriented along a lattice diagonal (a 45 degrees spiral) either three
or one hole pocket are populated. As we will see, the zero degree spiral is 
realized for intermediate values of $\rho_s$, while the 45 degrees spiral is 
unstable against the formation of inhomogeneities in the fermion density. 
Interesting calculations of a similar nature were presented in 
\cite{Sus04,Kot04} in the context of microscopic $t$-$J$-like models. In these 
works, in a particular parameter range a zero degree spiral has been identified
as the most stable 
configuration. Our effective field theory treatment complements these results 
in an interesting way. First, it is model-independent and thus applicable to a 
large variety of microscopic systems, because the most general form of the 
effective action is taken into account. In addition, it is controlled by a 
systematic low-energy expansion. Before we discuss homogeneous versus spiral 
phases, we address the issue of phase separation in the context of the $t$-$J$ 
model.

\subsection{Stability against Phase Separation}

It is well-known that the $t$-$J$ model shows phase separation for small values
of $t$. In this case the doped holes are heavy. Each hole is surrounded by four
bonds which, in the absence of the hole, would carry a negative 
antiferromagnetic contribution to the energy. In order to minimize the number 
of broken antiferromagnetic bonds, the holes may accumulate in some region of 
the lattice, thus leaving an otherwise undoped antiferromagnet behind, i.e.\
the system undergoes phase separation.

The energy density of the undoped antiferromagnet is $\epsilon_0$. In the 
$t$-$J$ model, which reduces to the Heisenberg model at half-filling, the 
energy density was determined with a very efficient loop-cluster algorithm as 
$\epsilon_0 = - 0.6693(1) J/a^2$, where $J$ is the exchange coupling and $a$ is
the lattice spacing \cite{Wie94}. A doped hole propagating in the 
antiferromagnet has mass $M$. Hence, to leading order in the fermion density 
$n$, the energy density of a doped antiferromagnet is 
\begin{equation}
\epsilon = \epsilon_0 + M n + {\cal O}(n^2).
\end{equation}
When the system phase separates, the doped holes accumulate in some region of
volume $V'$, leaving an otherwise undoped antiferromagnet behind in the 
remaining volume $V - V'$. If there are no electrons at all in the hole-rich
region, each hole occupies an area $a^2$ and hence $V' = N a^2$, where $N$ is 
the number of holes. In the $t$-$J$ model, the unoccupied region of volume $V'$
does not contribute to the energy density, while in the remaining volume 
$V - V'$ one has the energy density $\epsilon_0$ of an undoped 
antiferromagnet. Hence, the energy density of a phase separated system is
\begin{equation}
\epsilon_{\text{ps}} = \frac{\epsilon_0 (V - V')}{V} = \epsilon_0 (1 - a^2 n).
\end{equation}
The doped $t$-$J$ model is thus stable against phase separation as long as
\begin{equation}
\epsilon < \epsilon_{\text{ps}} \ \Rightarrow \ M < - \epsilon_0 a^2 = 
0.6693(1) J.
\end{equation}
In the $t$-$J$ model at small $t$ the holes are essentially static, breaking
four antiferromagnetic bonds, while $\epsilon_0$ represents the energy of two
intact antiferromagnetic bonds. Hence, for small $t$ one has
$M \approx - 2 \epsilon_0 a^2$ which implies phase separation. However, the 
hole mass $M$ decreases with increasing $t$ and can even become negative
\cite{Bru00}. Hence, for sufficiently large $t$, one indeed obtains a doped 
antiferromagnet which is stable against phase separation. It should be noted
that a negative rest mass $M$ is not at all problematical from a theoretical 
point of view. It only means that the doped antiferromagnet has a lower energy 
than the undoped system. Still, since particle number is conserved, the system 
cannot lower its energy by creating fermions.

\subsection{Fermionic Contribution to the Energy}

In this subsection we compute the fermionic contribution to the energy, keeping
the parameters $c^3_i$ and $c^\pm_i$ of the spiral fixed. For the moment, we
ignore the 4-fermion couplings. The considerations of this paper are valid only
if the 4-fermion couplings are weak and can be treated in perturbation theory.
Furthermore, we may neglect the hole-two-magnon vertices proportional to $N_1$ 
and $N_2$ which involve two spatial derivatives and are thus of higher order 
than the hole-one-magnon vertex proportional to $\Lambda$. The Euclidean action
of eq.(\ref{Lagrangian2}) then gives rise to the fermion Hamiltonian
\begin{align}
H=&\int d^2x \sum_{\ontopof{f=\alpha,\beta}{\, s = +,-}} 
\Big[ M \Psi^{f\dagger}_s \Psi^f_s +
\frac{1}{2 M'} D_i \Psi^{f\dagger}_s D_i \Psi^f_s \nonumber \\
&+\sigma_f \frac{1}{2 M''} (D_1 \Psi^{f\dagger}_s D_2 \Psi^f_s +
D_2 \Psi^{f\dagger}_s D_1 \Psi^f_s) + 
\Lambda (\Psi^{f\dagger}_s v^s_1 \Psi^f_{-s} 
+ \sigma_f \Psi^{f\dagger}_s v^s_2 \Psi^f_{-s})\Big],
\end{align}
with the covariant derivative
\begin{equation}
D_i \Psi^f_\pm(x) = \p_i \Psi^f_\pm(x) \pm i v^3_i(x) \Psi^f_\pm(x).
\end{equation}
Here $\Psi^{f\dagger}_s(x)$ and $\Psi^f_s(x)$ are creation and annihilation 
operators (not Grassmann numbers) for fermions of flavor $f = \alpha,\beta$ and
spin $s = +,-$ (parallel or antiparallel to the local staggered magnetization),
which obey canonical anticommutation relations. As before, $\sigma_\alpha = 1$ 
and $\sigma_\beta = - 1$. The above Hamiltonian is invariant against 
time-independent $U(1)_s$ gauge transformations
\begin{eqnarray}
&&\Psi^f_\pm(x)' = \exp(\pm i \alpha(x)) \Psi^f_\pm(x), \nonumber \\
&&{v^3_i}(x)' = v^3_i(x) - \p_i \alpha(x), \quad
{v^\pm_i}(x)' = v^\pm_i(x) \exp(\pm 2 i \alpha(x)).
\end{eqnarray}
Here we consider holes propagating in the background of a configuration with a
spiral in the staggered magnetization. Based on the considerations in the 
appendix, we can then limit ourselves to
\begin{equation}
{v^3_i}(x)' = c^3_i, \quad {v^\pm_i}(x)' = c_i \in \R.
\end{equation}
Hence, after the gauge transformation, the fermions experience a constant
composite vector field ${v_i}(x)'$. The Hamiltonian can then be diagonalized by
going to momentum space. Since magnon exchange does not mix the flavors, the
Hamiltonian can be considered separately for $f = \alpha$ and $f = \beta$, but
it still mixes spin $s = +$ with $s = -$. The single-particle Hamiltonian for 
holes with spatial momentum $\vec p = (p_1,p_2)$ takes the form
\begin{equation}
\label{Hf}
H^f(\vec p) = \left(\begin{array}{cc}
M + \frac{(p_i - c_i^3)^2}{2 M'} + 
\sigma_f \frac{(p_1 - c_1^3)(p_2 - c_2^3)}{M''} &
\Lambda (c_1 + \sigma_f c_2) \\ \Lambda (c_1 + \sigma_f c_2) &
M + \frac{(p_i + c_i^3)^2}{2 M'} + 
\sigma_f \frac{(p_1 + c_1^3)(p_2 + c_2^3)}{M''} \end{array} \right).
\end{equation}
The hole-one-magnon vertex proportional to $\Lambda$ mixes the spin $s = +$ and
$s = -$ states and provides a potential mechanism to stabilize a spiral phase.
The diagonalization of the above Hamiltonian yields
\begin{align}
\label{energy}
E^f_\pm(\vec p)=\,\, & M + \frac{p_i^2 + (c_i^3)^2}{2 M'} + 
\sigma_f \frac{p_1 p_2 + c_1^3 c_2^3}{M''} \nonumber \\
&\pm \sqrt{\left(\frac{p_i c_i^3}{M'} + 
\sigma_f \frac{p_1 c_2^3 + p_2 c_1^3}{M''}\right)^2 
+ \Lambda^2 (c_1 + \sigma_f c_2)^2}.
\end{align}
In particular, mixing via the $\Lambda$ vertex lowers the energy $E^f_-$ 
and raises the energy $E^f_+$. It should be noted that in this case the index 
$\pm$ no longer refers to the spin orientation. Indeed, the eigenvectors 
corresponding to $E^f_\pm$ are linear combinations of both spins. The minimum 
of the energy is located at $\vec p = 0$ for which
\begin{equation}
E^f_\pm(0) = M + \frac{(c_i^3)^2}{2 M'} + \sigma_f \frac{c_1^3 c_2^3}{M''}
\pm \Lambda |c_1 + \sigma_f c_2|.
\end{equation}
Since $c_i^3$ does not affect the magnon contribution to the energy density, we
fix it by minimizing $E^f_-(0)$ which implies $c_1^3 = c_2^3 = 0$. According to
eq.(\ref{constant}) this implies that $\theta_0 = \frac{\pi}{2}$, i.e.\ the 
spiral is along a great circle on the sphere $S^2$. By repeating the whole
calculation including terms of ${\cal O}((c_i^3)^2)$, we have verified a 
posteriori that putting $c_i^3 = 0$ is indeed justified. At present we cannot
exclude that there might be a minimum with a lower energy for large values of 
the $c_i^3$. Investigating this issue would require a somewhat tedious 
numerical calculation which we have not yet performed. For $c_1^3 = c_2^3 = 0$ 
the energies of eq.(\ref{energy}) reduce to
\begin{equation}
E^f_\pm(\vec p) = M + \frac{p_i^2}{2 M'} + \sigma_f \frac{p_1 p_2}{M''} \pm
\Lambda |c_1 + \sigma_f c_2|.
\end{equation}
Consequently, the filled hole pockets $P^f_\pm$ (with $M' < M''$) are ellipses 
determined by
\begin{equation}
\frac{p_i^2}{2 M'} + \sigma_f \frac{p_1 p_2}{M''} = T^f_\pm,
\end{equation}
where $T^f_\pm$ is the kinetic energy of a hole in the pocket $P^f_\pm$ at the 
Fermi surface. The area of an occupied hole pocket determines the fermion 
density as
\begin{equation}
n^f_\pm = \frac{1}{(2 \pi)^2} \int_{P^f_\pm} d^2p = 
\frac{1}{2 \pi} M_\eff T^f_\pm, \quad
M_\eff = \frac{M' M''}{\sqrt{{M''}^2 - {M'}^2}}.
\end{equation}
The kinetic energy density of a filled pocket is given by
\begin{equation}
t^f_\pm = \frac{1}{(2 \pi)^2} \int_{P^f_\pm} d^2p 
\left(\frac{p_i^2}{2 M'} + \sigma_f \frac{p_1 p_2}{M''}\right) =
\frac{1}{4 \pi} M_\eff {T^f_\pm}^2.
\end{equation}
The total density of fermions of all flavors is
\begin{equation}
n = n^\alpha_+ + n^\alpha_- + n^\beta_+ + n^\beta_- = \frac{1}{2 \pi}
M_\eff(T^\alpha_+ + T^\alpha_- + T^\beta_+ + T^\beta_-),
\end{equation}
and the total energy density of the holes is
\begin{equation}
\epsilon_h = \epsilon^\alpha_+ + \epsilon^\alpha_- + 
\epsilon^\beta_+ + \epsilon^\beta_-,
\end{equation}
with
\begin{equation}
\epsilon^f_\pm = (M \pm \Lambda |c_1 + \sigma_f c_2|) n^f_\pm + t^f_\pm.
\end{equation}
The filling of the various hole pockets is controlled by the parameters
$T^f_\pm$ which must be varied in order to minimize the energy while keeping 
the total density of holes fixed. We thus introduce
\begin{equation}
S = \epsilon_h - \mu n,
\end{equation}
where $\mu$ is a Lagrange multiplier that fixes the density, and we demand
\begin{equation}
\label{mini}
\frac{\p S}{\p T^f_\pm} = 
\frac{1}{2 \pi} M_\eff(M \pm \Lambda |c_1 + \sigma_f c_2| + 
T^f_\pm - \mu) = 0.
\end{equation}
One may wonder if the density of holes of flavor 
$\alpha$ and $\beta$ should not be fixed separately. After all, flavor is
conserved due to the accidental $U(1)_F$ symmetry. While the $U(1)_F$ symmetry 
arises for the leading terms in the effective action, it is not present at the
microscopic level. Although they enter the effective theory only at higher 
orders, there are physical processes that can change flavor. Despite the fact
that such processes are rare, it would hence be inappropriate to fix the 
fermion numbers of different flavors separately.

\subsection{Four Populated Hole Pockets: Homogeneous Phase}

We will now populate the various hole pockets with fermions. First, we keep the
configuration of the staggered magnetization fixed and we vary the $T^f_\pm$ in
order to minimize the energy of the fermions. Then we also vary the parameters 
$c_i$ of the magnon field in order to minimize the total energy. One must 
distinguish various cases, depending on how many hole pockets are populated 
with fermions. In this subsection, we consider the case of populating all four 
hole pockets (i.e.\ with both flavors $f = \alpha, \beta$ and with both energy
indices $\pm$). In this case, eq.(\ref{mini}) implies
\begin{equation}
\label{Teff}
\mu = M + \frac{\pi n}{2 M_\eff}, \quad 
T^f_\pm = \frac{\pi n}{2 M_\eff} \mp \Lambda |c_1 + \sigma_f c_2|.
\end{equation}
The total energy density then takes the form
\begin{eqnarray}
\epsilon&=&\epsilon_0 + \epsilon_m + \epsilon_h = \epsilon_0 + 
2 \rho_s (c_1^2 + c_2^2) + 
\epsilon^\alpha_+ + \epsilon^\alpha_- + \epsilon^\beta_+ + \epsilon^\beta_-
\nonumber \\
&=&\epsilon_0 + 2 \rho_s (c_1^2 + c_2^2) + M n + 
\frac{\pi n^2}{4 M_\eff} - 
\frac{1}{\pi} M_\eff \Lambda^2 (c_1^2 + c_2^2).
\end{eqnarray}
For $2 \pi \rho_s > M_\eff \Lambda^2$ the energy is minimized for $c_i = 0$
and the configuration is thus homogeneous. The total energy density in the 
four-pocket case is then given by
\begin{equation}
\label{etothom}
\epsilon_4 = \epsilon_0 + M n + \frac{\pi n^2}{4 M_\eff}.
\end{equation}
The homogeneous phase is shown in figure 2.
\begin{figure}[t]
\begin{center}
\epsfig{file=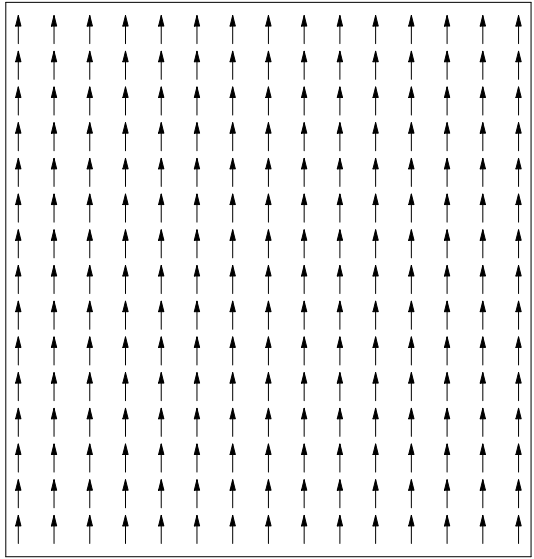,width=7cm}
\end{center}
\caption{\it The homogeneous phase with constant staggered magnetization.}
\end{figure}
For $2 \pi \rho_s < M_\eff \Lambda^2$, on the other hand, the energy is not 
bounded from below. In this case, $c_1^2 + c_2^2$ seems to grow without bound.
However, according to eq.(\ref{Teff}) this would lead to $T^f_+ < 0$ which is
physically meaningless. What really happens is that one pocket gets completely
emptied and we must thus turn to the three-pocket case.

\subsection{Three Populated Hole Pockets: 45 Degrees Spiral}

We now populate only three pockets with holes: the two pockets with the lower 
energies $E^\alpha_-$ and $E^\beta_-$ as well as the pocket with the higher 
energy $E^\alpha_+$. Of course, alternatively one could also fill the 
$\beta_+$-pocket. We now obtain
\begin{equation}
n = n^\alpha_+ + n^\alpha_- + n^\beta_- = 
\frac{1}{2 \pi} M_\eff(T^\alpha_+ + T^\alpha_- + T^\beta_-), 
\quad \epsilon_h = \epsilon^\alpha_+ + \epsilon^\alpha_- + \epsilon^\beta_-,
\end{equation}
such that eq.(\ref{mini}) yields
\begin{eqnarray}
&&\mu = M + \frac{2 \pi n}{3 M_\eff} - 
\frac{\Lambda}{3} |c_1 - c_2|, \nonumber \\
&&T^\alpha_+ = \frac{2 \pi n}{3 M_\eff} - \Lambda |c_1 + c_2| -
\frac{\Lambda}{3} |c_1 - c_2|, \nonumber \\
&&T^\alpha_- = \frac{2 \pi n}{3 M_\eff} + \Lambda |c_1 + c_2| -
\frac{\Lambda}{3} |c_1 - c_2|, \nonumber \\
&&T^\beta_- = \frac{2 \pi n}{3 M_\eff} +
\frac{2 \Lambda}{3} |c_1 - c_2|.
\end{eqnarray}
The total energy density then takes the form
\begin{eqnarray}
\label{e3}
\epsilon&=&\epsilon_0 + \epsilon_m + \epsilon_h = \epsilon_0 +
2 \rho_s (c_1^2 + c_2^2) + 
\epsilon^\alpha_+ + \epsilon^\alpha_- + \epsilon^\beta_- \nonumber \\
&=&\epsilon_0 + 2 \rho_s (c_1^2 + c_2^2) + 
\left(M - \frac{\Lambda}{3} |c_1 - c_2|\right) n + 
\frac{\pi n^2}{3 M_\eff} \nonumber \\
&&-\,\, \frac{2}{3 \pi} M_\eff \Lambda^2 (c_1^2 + c_1 c_2 + c_2^2).
\end{eqnarray}
For $2 \pi \rho_s > M_\eff \Lambda^2$ the energy density is bounded from 
below and its minimum is located at $c_1 = - c_2$ with
\begin{equation}
|c_1| = |c_2| = \frac{\pi}{2} 
\frac{\Lambda n}{6 \pi \rho_s - M_\eff \Lambda^2}.
\end{equation}
This represents a spiral in the staggered magnetization oriented along a 
lattice diagonal --- a 45 degrees spiral. When one occupies the $\beta_+$-
instead of the $\alpha_+$-pocket, one finds $c_1 = c_2$, i.e.\ the spiral is 
then oriented in the orthogonal diagonal direction. According to 
eq.(\ref{pitch}) in the appendix, the spiral pitch (i.e.\ the wave number of 
the spiral) is given by
\begin{equation}
k = 2 \sqrt{c_i^3 c_i^3 + c_i c_i} = 2 \sqrt{c_1^2 + c_2^2} =
\frac{\sqrt{2} \pi \Lambda n}{6 \pi \rho_s - M_\eff \Lambda^2}.
\end{equation}
The resulting energy density in the three-pocket case takes the form
\begin{equation}
\label{etot3}
\epsilon_3 = \epsilon_0 + M n + 
\frac{\pi}{3 M_\eff} \left(1 - \frac{1}{2} \frac{M_\eff \Lambda^2}
{6 \pi \rho_s - M_\eff \Lambda^2}\right) n^2.
\end{equation}
The 45 degrees spiral is illustrated in figure 3.
\begin{figure}[t]
\begin{center}
\epsfig{file=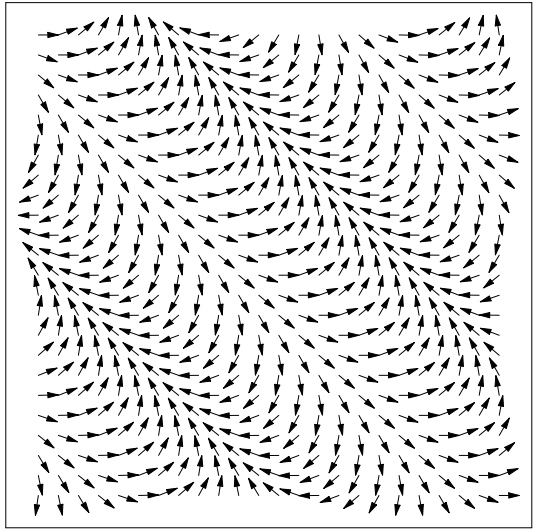,width=7cm}
\end{center}
\caption{\it A 45 degrees spiral oriented along a lattice diagonal.}
\end{figure}
It is energetically less favorable than the homogeneous phase because 
$\epsilon_3 > \epsilon_4$ for $2 \pi \rho_s > M_\eff \Lambda^2$. Only for 
$2 \pi \rho_s = M_\eff \Lambda^2$ both phases cost the same energy, i.e.\
$\epsilon_3 = \epsilon_4$. For $2 \pi \rho_s < M_\eff \Lambda^2$ the energy 
density of eq.(\ref{e3}) is unbounded from below and $c_1^2 + c_2^2$ again 
seems to grow without bound. This, however, would lead to $T^\alpha_+ < 0$ 
which is unphysical. In fact, the $\alpha_+$-pocket is completely emptied and 
we are thus led to investigate the two-pocket case.

\subsection{Two Populated Hole Pockets: Zero Degree Spiral}

We now populate only two pockets with holes. These are necessarily the pockets
with the lower energies $E^\alpha_-$ and $E^\beta_-$. In this case we have
\begin{equation}
n = n^\alpha_- + n^\beta_- = \frac{1}{2 \pi} M_\eff(T^\alpha_- + T^\beta_-), 
\quad \epsilon_h = \epsilon^\alpha_- + \epsilon^\beta_-,
\end{equation}
and thus eq.(\ref{mini}) now implies
\begin{eqnarray}
&&\mu = M + \frac{\pi n}{M_\eff} - 
\frac{\Lambda}{2} (|c_1 + c_2| + |c_1 - c_2|), \nonumber \\
&&T^\alpha_- = \frac{\pi n}{M_\eff} + 
\frac{\Lambda}{2}(|c_1 + c_2| - |c_1 - c_2|), \nonumber \\
&&T^\beta_- = \frac{\pi n}{M_\eff} + 
\frac{\Lambda}{2}(|c_1 - c_2| - |c_1 + c_2|).
\end{eqnarray}
The total energy density then takes the form
\begin{eqnarray}
\label{e0degree}
\epsilon&=&\epsilon_0 + \epsilon_m + \epsilon_h = \epsilon_0 +
2 \rho_s (c_1^2 + c_2^2) + \epsilon^\alpha_- + \epsilon^\beta_- 
\nonumber \\
&=&\epsilon_0 + 2 \rho_s (c_1^2 + c_2^2) + 
\left(M - \frac{\Lambda}{2}(|c_1 + c_2| + |c_1 - c_2|)\right) n + 
\frac{\pi n^2}{2 M_\eff} \nonumber \\
&&-\,\,\frac{1}{2 \pi} M_\eff \frac{\Lambda^2}{4} (|c_1 + c_2| - |c_1 - c_2|)^2.
\end{eqnarray}
For $2 \pi \rho_s > \frac{1}{2} M_\eff \Lambda^2$ the energy density is
bounded from below and has its minimum at
\begin{equation}
|c_1| = \frac{\Lambda}{4 \rho_s} n, \ |c_2| = 0, \quad \mbox{or} \
|c_1| = 0, \ |c_2| = \frac{\Lambda}{4 \rho_s} n,
\end{equation}
which corresponds to a spiral in the staggered magnetization oriented along a 
lattice axis --- a zero degree spiral. Again, according to eq.(\ref{pitch}), 
the wave number of the spiral is given by
\begin{equation}
k = 2 \sqrt{c_i^3 c_i^3 + c_i c_i} = 2 \sqrt{c_1^2 + c_2^2} =
\frac{\Lambda n}{2 \rho_s},
\end{equation}
and the resulting energy density in the two-pocket case takes the form
\begin{equation}
\label{etotzero}
\epsilon_2 = \epsilon_0 + M n + 
\left(\frac{\pi}{2 M_\eff} - \frac{\Lambda^2}{8 \rho_s}\right) n^2.
\end{equation}
The zero degree spiral is more stable than the homogeneous phase if 
$\epsilon_2 < \epsilon_4$, which is the case for 
$2 \pi \rho_s < M_\eff \Lambda^2$. This also clarifies the instability of the
homogeneous phase in this parameter regime: it simply turns into the zero 
degree spiral. We have verified explicitly that the assumption $c_i^3 = 0$ is 
justified by repeating the whole calculation including terms up to 
${\cal O}((c_i^3)^2)$. Indeed, $c_i^3 = 0$ turns out to be a stable minimum for
$2 \pi \rho_s < M_\eff \Lambda^2$. The zero degree spiral is illustrated in 
figure 4.
\begin{figure}[t]
\begin{center}
\epsfig{file=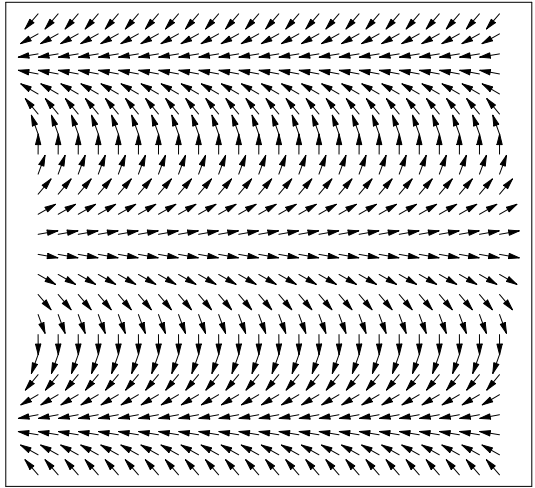,width=7cm}
\end{center}
\caption{\it A zero degree spiral oriented along a lattice axis.}
\end{figure}

\subsection{One Populated Hole Pocket: 45 Degrees Spiral}

Finally, let us populate only one hole pocket, say the states with energy 
$E^\alpha_-$. Of course, alternatively one could also occupy the 
$\beta_-$-pocket. One now obtains
\begin{equation}
T^\alpha_- = \frac{2 \pi n}{M_\eff}.
\end{equation}
The total energy density then takes the form
\begin{eqnarray}
\epsilon&=&\epsilon_0 + \epsilon_m + \epsilon_h = \epsilon_0 +
2 \rho_s (c_1^2 + c_2^2) + \epsilon^\alpha_- \nonumber \\
&=&\epsilon_0 + 2 \rho_s (c_1^2 + c_2^2) + 
(M - \Lambda |c_1 + c_2|) n + \frac{\pi n^2}{M_\eff},
\end{eqnarray}
which is minimized for $c_1 = c_2$ with
\begin{equation}
|c_1| = |c_2| = \frac{\Lambda}{4 \rho_s} n.
\end{equation}
This again represents a 45 degrees spiral in the staggered magnetization 
oriented along a lattice diagonal. When one occupies the $\beta_-$-pocket one
finds $c_1 = - c_2$, i.e., as in the three-pocket case, the $\alpha$- and 
$\beta$-spirals are oriented in orthogonal diagonal directions. Note that the 
three-pocket 45 degrees spiral with an occupied $\alpha_+$-pocket is
oriented in a direction orthogonal to the one of the one-pocket 45 degrees 
spiral with an occupied $\alpha_-$-pocket. In the one-pocket case, the spiral 
pitch is given by
\begin{equation} 
k = 2 \sqrt{c_i^3 c_i^3 + c_i c_i} = 2 \sqrt{c_1^2 + c_2^2} =
\frac{\Lambda n}{\sqrt{2} \rho_s},
\end{equation}
and the corresponding energy density takes the form
\begin{equation}
\label{etot45}
\epsilon_1 = \epsilon_0 + M n + 
\left(\frac{\pi}{M_\eff} - \frac{\Lambda^2}{4 \rho_s}\right) n^2.
\end{equation}
The 45 degrees spiral is energetically more favorable than the zero degree 
spiral if $\epsilon_1 < \epsilon_2$, which is the case for 
$2 \pi \rho_s < \frac{1}{2} M_\eff \Lambda^2$.  Again, by repeating the whole
calculation including terms up to ${\cal O}((c_i^3)^2)$, we have verified 
explicitly that for $2 \pi \rho_s < M_\eff \Lambda^2$ the assumption of
$c_i^3 = 0$ is justified because it corresponds to a stable minimum.

However, as already pointed out in \cite{Aue91,Arr91,Chu95,Sus04,Kot04}, the 
negative coefficient of the term proportional to $n^2$ (i.e.\ a negative
compressibility) leads to an instability of the 45 degrees spiral against 
increasing the local fermion density by decreasing the wavelength $2 \pi/k$ of 
the spiral. To see this, let us consider a one-pocket 45 degrees spiral 
filling a 
sub-volume $V'$ and leaving an undoped antiferromagnet behind in the remaining 
volume $V - V'$. The energy density of this configuration with an inhomogeneous
fermion density is given by
\begin{eqnarray}
\epsilon_{\text{inh}}&=&\epsilon_1 \frac{V'}{V} + 
\epsilon_0 \frac{V - V'}{V} = \epsilon_0 + 
\left[M n + \left(\frac{\pi}{M_\eff} - \frac{\Lambda^2}{4 \rho_s}\right) n^2
\right] \frac{V'}{V} \nonumber \\
&=&\epsilon_0 + \left[M \frac{N}{V'} + 
\left(\frac{\pi}{M_\eff} - \frac{\Lambda^2}{4 \rho_s}\right) 
\left(\frac{N}{V'}\right)^2\right] \frac{V'}{V} \nonumber \\
&=&\epsilon_0 + \left[M +
\left(\frac{\pi}{M_\eff} - \frac{\Lambda^2}{4 \rho_s}\right) \frac{N}{V'}
\right] \frac{N}{V}.
\end{eqnarray}
Since the coefficient of the term proportional to $1/V'$ is negative for
$2 \pi \rho_s < \frac{1}{2} M_\eff \Lambda^2$, the system can minimize its 
energy by shrinking $V'$ and thus increasing the fermion density $n$ in the 
region of the 45 degrees spiral. The one-pocket 45 degrees spiral is thus 
unstable against the formation of inhomogeneities in the fermion density. Our 
basic assumption that the system is homogeneous is then not satisfied.

\subsection{Symmetry Breaking Pattern in the Spiral Phase}

As we have seen, for intermediate values of $\rho_s$ a spiral phase replaces 
the phase with homogeneous staggered magnetization. In the homogeneous phase 
the $SU(2)_s$ spin symmetry is spontaneously broken to its $U(1)_s$ subgroup. 
Due to antiferromagnetism, the displacement by one lattice spacing is also 
spontaneously broken. In a spiral phase, on the other hand, a $U(1)_s$ spin 
rotation that leaves the configuration invariant no longer exists. Hence, 
$U(1)_s$ is then also spontaneously broken and one may expect an additional 
massless Goldstone boson. Furthermore, due to the finite wavelength 
$2 \pi/k$ of the spiral, translation invariance (not 
only by one lattice spacing) is also spontaneously broken. Only the 
translations by an integer multiple of the spiral wavelength remain unbroken. 
Due to the spontaneously broken translation symmetry one expects yet another
massless Goldstone boson --- a spiral phonon (or helimagnon) --- which 
corresponds to vibrations of the spiral. In order to correctly count the number
of Goldstone bosons one must notice that while both $U(1)_s$ and translation 
invariance are individually spontaneously broken, there is a combination of 
these two symmetries that remains unbroken. In particular, any translation of 
the spiral configuration can be compensated by an appropriate $U(1)_s$ spin 
rotation. Consequently, there are not two but there is only one additional 
Goldstone boson --- the spiral phonon. Besides this additional massless boson, 
in the spiral configuration there are still two magnons --- a transverse and a 
longitudinal one. The transverse magnon corresponds to spin fluctuations out of
the spiral plane, while the longitudinal magnon represents in-plane 
fluctuations. 

It should be noted that the wavelength $2 \pi/k \propto 1/n$ 
of the spiral represents a new length scale in the problem. In particular, for 
very small fermion density $n$ this length scale is arbitrarily long. In that 
case, the spiral phonon will have very little effect on the dynamics of the 
holes. When the fermion density increases, the spiral wavelength shrinks and 
the spiral phonon becomes more important. In particular, besides magnon 
exchange the exchange of spiral phonons may then contribute significantly to 
the long-range interactions between the holes.  

Since the symmetry is no longer broken just from $SU(2)_s$ to $U(1)_s$, one may
wonder whether the vector $\vec e(x)$ living in the coset space 
$SU(2)_s/U(1)_s = S^2$ is still an appropriate low-energy degree of freedom in
the spiral phase. Fortunately, this is indeed the case. Since the spiral phonon
arises as an additional Goldstone boson, one may also wonder whether a new 
field must be added to the effective Lagrangian. This is not necessary because,
just like the two magnons, the spiral phonon arises as a fluctuation of the 
staggered magnetization vector $\vec e(x)$ (in this case around the spiral 
configuration). Only if one would construct another effective theory valid
only at extremely long length scales much larger than the wavelength of the
spiral, the basic fields of the theory would have to be redefined. In such an 
effective theory the spiral itself would be a short distance phenomenon and the
spiral phonon would appear explicitly as an independent degree of freedom. The 
corresponding physics at extremely low energies is still contained in our 
effective theory which is valid at length scales both longer and shorter than 
the wavelength of the spiral.

\section{Inclusion of 4-Fermion Couplings in \\ Perturbation Theory}

In this section the 4-fermion contact interactions are incorporated in 
perturbation theory. Depending on the microscopic system in question, the
4-fermion couplings may or may not be small. If they are large, the result of
the perturbative calculation should not be trusted. In that case, one could
still perform a variational calculation. In this work we limit ourselves to
first order perturbation theory. We will distinguish four cases: the 
homogeneous phase, the three-pocket 45 degrees spiral, the zero degree spiral, 
and the one-pocket 45 degrees spiral. Finally, depending on the values of the 
low-energy constants, we determine which phase is energetically favorable.

\subsection{Four-Pocket Case: Homogeneous Phase}

Let us first consider the homogeneous phase. The perturbation of the 
Hamiltonian due to the leading 4-fermion contact terms of 
eq.(\ref{Lagrange4}) is given by
\begin{align}
\label{DeltaH}
\Delta H = \int d^2x \sum_{s = +,-} \Big[ &
\frac{G_1}{2} (\Psi^{\alpha\dagger}_s \Psi^\alpha_s 
\Psi^{\alpha\dagger}_{-s} \Psi^\alpha_{-s} + 
\Psi^{\beta\dagger}_s \Psi^\beta_s 
\Psi^{\beta\dagger}_{-s} \Psi^\beta_{-s}) \nonumber \\[-1ex]
&+G_2 \Psi^{\alpha\dagger}_s \Psi^\alpha_s \Psi^{\beta\dagger}_s \Psi^\beta_s +
G_3 \Psi^{\alpha\dagger}_s \Psi^\alpha_s 
\Psi^{\beta\dagger}_{-s} \Psi^\beta_{-s} \Big].
\end{align}
It should be noted that $\Psi^{f\dagger}_s(x)$ and $\Psi^f_s(x)$ again are 
fermion creation and annihilation operators (and not Grassmann numbers).  
The term proportional to $G_4$ is of higher order and hence need not be
considered here. In the homogeneous phase the fermion density is equally 
distributed among the two spin orientations and the two flavors such that
\begin{equation}
\langle \Psi^{\alpha\dagger}_+ \Psi^\alpha_+ \rangle =
\langle \Psi^{\alpha\dagger}_- \Psi^\alpha_- \rangle =
\langle \Psi^{\beta\dagger}_+ \Psi^\beta_+ \rangle =
\langle \Psi^{\beta\dagger}_- \Psi^\beta_- \rangle = \frac{n}{4}.
\end{equation}
The brackets denote expectation values in the unperturbed state determined in
section 4.3. Since the fermions are uncorrelated, for $f \neq f'$ or 
$s \neq s'$ one has
\begin{equation}
\langle \Psi^{f\dagger}_s \Psi^f_s \Psi^{f'\dagger}_{s'} \Psi^{f'}_{s'} \rangle
= \langle \Psi^{f\dagger}_s \Psi^f_s \rangle 
\langle \Psi^{f'\dagger}_{s'} \Psi^{f'}_{s'}\rangle.
\end{equation}
Taking the 4-fermion contact terms into account in first order perturbation
theory, the total energy density of eq.(\ref{etothom}) receives an additional 
contribution and now reads
\begin{equation}
\label{eps4}
\epsilon_4 = \epsilon_0 + M n + \frac{\pi n^2}{4 M_\eff} +
\frac{1}{8} (G_1 + G_2 + G_3) n^2.
\end{equation}

\subsection{Three-Pocket Case: 45 Degrees Spiral}

For $c^3_i = 0$ the eigenvectors of the single-particle Hamiltonian of 
eq.(\ref{Hf}) corresponding to the energy eigenvalues $E^f_\pm(\vec p)$ are 
given by
\begin{equation}
\widetilde \Psi^f_\pm = \frac{1}{\sqrt{2}}(\Psi^f_- \pm \Psi^f_+) \
\Rightarrow \
\Psi^f_\pm = \frac{1}{\sqrt{2}}(\widetilde \Psi^f_+ \mp \widetilde \Psi^f_-).
\end{equation}
Inserting this expression in eq.(\ref{DeltaH}) allows us to evaluate the
expectation value $\langle \Delta H \rangle$ in the unperturbed states
determined in section 4. In the three-pocket case the states 
with energies $E^\alpha_-(\vec p)$, $E^\beta_-(\vec p)$, as well as 
$E^\alpha_+(\vec p)$ (or alternatively $E^\beta_+(\vec p)$), and with $\vec p$ 
inside the respective hole pocket are occupied and
\begin{eqnarray}
&&\langle \widetilde \Psi^{\alpha\dagger}_+ \widetilde \Psi^\alpha_+ \rangle =
\langle \widetilde \Psi^{\alpha\dagger}_- \widetilde \Psi^\alpha_- \rangle =
\left(1 - \frac{1}{2} \frac{M_\eff \Lambda^2}
{6 \pi \rho_s - M_\eff \Lambda^2}\right) \frac{n}{3}, \nonumber \\
&&\langle \widetilde \Psi^{\beta\dagger}_- \widetilde \Psi^\beta_- \rangle =
\left(1 + \frac{M_\eff \Lambda^2}{6 \pi \rho_s - M_\eff \Lambda^2}\right) 
\frac{n}{3}, \quad
\langle \widetilde \Psi^{\beta\dagger}_+ \widetilde \Psi^\beta_+ \rangle = 0.
\end{eqnarray}
As a result the energy density of eq.(\ref{etot3}) turns into
\begin{eqnarray}
\label{eps3}
\epsilon_3&=&\epsilon_0 + M n + 
\frac{\pi}{3 M_\eff}\left(1 - \frac{1}{2}
\frac{M_\eff \Lambda^2}{6 \pi \rho_s - M_\eff \Lambda^2}\right) n^2
\nonumber \\
&&+\,\, \frac{1}{4} 
\left[(G_1 + G_2 + G_3) 4 \pi \rho_s - G_1 M_\eff \Lambda^2 \right] 
\frac{4 \pi \rho_s - M_\eff \Lambda^2}
{(6 \pi \rho_s - M_\eff \Lambda^2)^2} n^2.
\end{eqnarray}

\subsection{Two-Pocket Case: Zero Degree Spiral}

In the zero degree spiral only the states 
with energy $E^\alpha_-(\vec p)$ and $E^\beta_-(\vec p)$ with $\vec p$ inside 
the respective hole pocket $P^f_-$ are occupied and hence
\begin{equation}
\langle \widetilde \Psi^{\alpha\dagger}_- \widetilde \Psi^\alpha_- \rangle =
\langle \widetilde \Psi^{\beta\dagger}_- \widetilde \Psi^\beta_- \rangle =
\frac{n}{2}, \quad
\langle \widetilde \Psi^{\alpha\dagger}_+ \widetilde \Psi^\alpha_+ \rangle =
\langle \widetilde \Psi^{\beta\dagger}_+ \widetilde \Psi^\beta_+ \rangle = 0.
\end{equation}
As a result the energy density of eq.(\ref{etotzero}) turns into
\begin{equation}
\label{eps2}
\epsilon_2 = \epsilon_0 + M n + \left(\frac{\pi}{2 M_\eff} - 
\frac{\Lambda^2}{8 \rho_s}\right) n^2 + \frac{1}{8} (G_2 + G_3) n^2.
\end{equation}

\subsection{One-Pocket Case: 45 Degrees Spiral}

In the one-pocket case only the states with energy $E^\alpha_-(\vec p)$ 
(or alternatively with $E^\beta_-(\vec p)$) and with $\vec p$ inside the
corresponding hole pocket are occupied such that
\begin{equation}
\langle \widetilde \Psi^{\alpha\dagger}_- \widetilde \Psi^\alpha_- \rangle = n,
\quad 
\langle \widetilde \Psi^{\alpha\dagger}_+ \widetilde \Psi^\alpha_+ \rangle =
\langle \widetilde \Psi^{\beta\dagger}_+ \widetilde \Psi^\beta_+ \rangle = 
\langle \widetilde \Psi^{\beta\dagger}_- \widetilde \Psi^\beta_- \rangle = 0.
\end{equation}
In this case the 4-fermion terms do not contribute to the energy density which
thus maintains the form of eq.(\ref{etot45}), i.e.
\begin{equation}
\label{eps1}
\epsilon_1 = \epsilon_0 + M n + 
\left(\frac{\pi}{M_\eff} - \frac{\Lambda^2}{4 \rho_s}\right) n^2.
\end{equation}

\subsection{Stability Ranges of Various Phases}

Let us summarize the results of the previous subsections. The energy densities 
of the various phases take the form
\begin{equation}
\epsilon_i = \epsilon_0 + M n + \frac{1}{2} \kappa_i n^2.
\end{equation}
According to eqs.(\ref{eps1}), (\ref{eps2}), (\ref{eps3}), and (\ref{eps4}), 
the compressibilities are given by
\begin{eqnarray}
\kappa_1&=&\frac{2 \pi}{M_\eff} - \frac{\Lambda^2}{2 \rho_s}, \nonumber \\
\kappa_2&=&\frac{\pi}{M_\eff} - \frac{\Lambda^2}{4 \rho_s}
+ \frac{1}{4} (G_2 + G_3), \nonumber \\
\kappa_3&=&\frac{2 \pi}{3 M_\eff}
\left(1 - \frac{1}{2} \frac{M_\eff \Lambda^2}
{6 \pi \rho_s - M_\eff \Lambda^2}\right) \nonumber \\
&&+\,\, \frac{1}{2} \left[(G_1 + G_2 + G_3) 4 \pi \rho_s - G_1 M_\eff \Lambda^2
\right] \frac{4 \pi \rho_s - M_\eff \Lambda^2}
{(6 \pi \rho_s - M_\eff \Lambda^2)^2}, \nonumber \\
\kappa_4&=&\frac{\pi}{2 M_\eff} + \frac{1}{4} (G_1 + G_2 + G_3).
\end{eqnarray}
The compressibilities $\kappa_i$ as functions of 
$M_\eff \Lambda^2/2 \pi \rho_s$ are shown in figure 5.
\begin{figure}[t]
\begin{center}
\epsfig{file=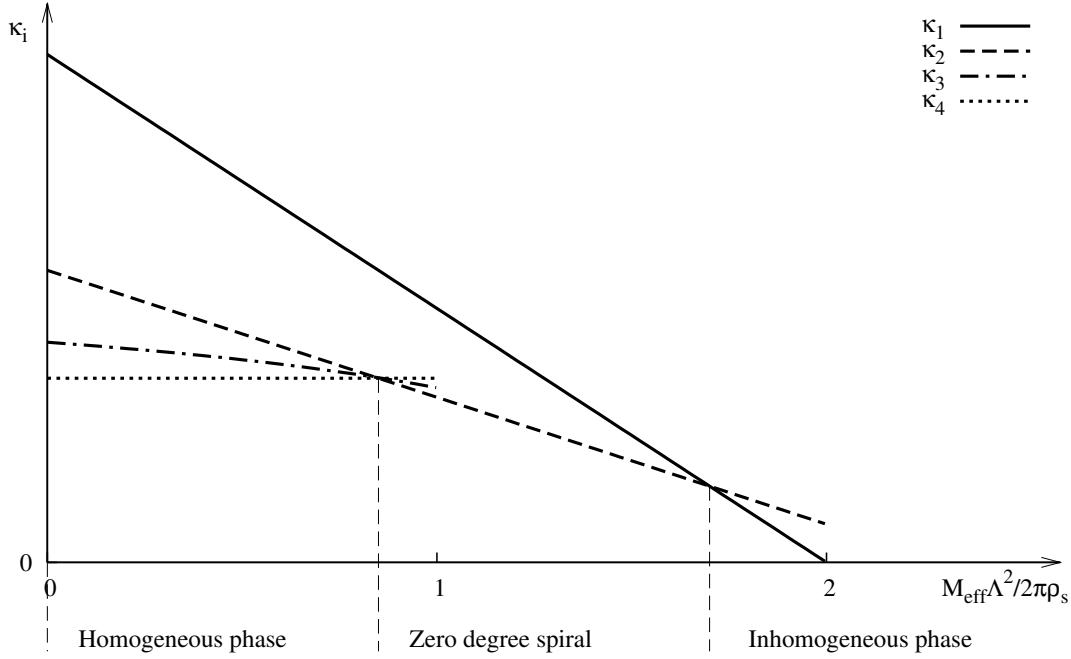,width=15cm}
\vspace{-2cm}
\end{center}
\caption{\it The compressibilities $\kappa_i$ (in the presence of weak 
repulsive 4-fermion couplings) as functions of 
$M_\eff \Lambda^2/2 \pi \rho_s$ determine the stability ranges of the various 
phases. A homogeneous phase, a zero degree spiral, or an inhomogeneous phase 
are energetically favorable, for large, intermediate, and small values of 
$\rho_s$, respectively. Only at the two isolated points separating the three 
regimes a 45 degrees spiral, either with one or with three populated hole
pockets, is degenerate with the other phases. The three- and four-pocket
cases become unstable at $M_\eff \Lambda^2/2 \pi \rho_s = 1$, and the one- and 
two-pocket cases become unstable at $M_\eff \Lambda^2/2 \pi \rho_s = 2$.}
\end{figure}
For large values of $\rho_s$, spiral phases cost a large amount of magnetic 
energy and a homogeneous phase is more stable. The energy density of the 
homogeneous phase is lower than the one of the spiral phases for 
$\kappa_4 < \kappa_1, \kappa_2, \kappa_3$, which is the case for
\begin{equation}
M_\eff \Lambda^2 + G_1 \frac{M_\eff^2 \Lambda^2}{2 \pi} \leq 2 \pi \rho_s 
\ \left\{\begin{array}{l} \mbox{Four-pocket case:} \\
\mbox{Homogeneous phase.} \end{array} \right.
\end{equation}
At ${\cal O}(G_i)$ the three-pocket 45 degrees spiral always costs more energy
than the other phases, except at the specific point
\begin{equation}
2 \pi \rho_s = M_\eff \Lambda^2 + G_1 \frac{M_\eff^2 \Lambda^2}{2 \pi}
\ \left\{\begin{array}{l} \mbox{Three-pocket case:} \\
\mbox{45 degrees spiral.} \end{array} \right.\
\end{equation}
For this particular value of $\rho_s$ the three-pocket 45 degrees spiral costs 
the same energy as the homogeneous phase and the zero degree spiral. For larger
values of the 4-fermion couplings (not accessible to first order perturbation 
theory) it is conceivable that the 45 degrees spiral becomes energetically 
favorable. This could be investigated, for example, using a variational 
calculation. For intermediate values of $\rho_s$ the homogeneous phase becomes 
unstable against the formation of a zero degree spiral, i.e.\ 
$\kappa_2 < \kappa_4$, and we find
\begin{equation}
 \frac{1}{2} M_\eff \Lambda^2 + 
(G_2 + G_3) \frac{M_\eff^2 \Lambda^2}{8 \pi} \leq 2 \pi \rho_s \leq 
M_\eff \Lambda^2 + 
G_1 \frac{M_\eff^2 \Lambda^2}{2 \pi}
\ \left\{\begin{array}{l} \mbox{Two-pocket case:} \\
\mbox{Zero degree spiral.} \end{array} \right. 
\end{equation}
For smaller values of $\rho_s$ the one-pocket 45 degrees spiral becomes 
energetically more favorable than the zero degree spiral, i.e.\ 
$\kappa_1 < \kappa_2$, but the one-pocket 45 degrees spiral exists only at the 
isolated point
\begin{equation}
2 \pi \rho_s = \frac{1}{2} M_\eff \Lambda^2 + 
(G_2 +G_3) \frac{M_\eff^2 \Lambda^2}{8 \pi} 
\ \left\{\begin{array}{l} \mbox{One-pocket case:} \\ 
\mbox{45 degrees spiral.} \end{array} \right.
\end{equation}
It may seem that a repulsive 4-fermion interaction $G_2 + G_3 > 0$ stabilizes 
the 45 degrees spiral, at least in the narrow parameter range down to
$2 \pi \rho_s =  \frac{1}{2} M_\eff \Lambda^2$. However, although we have not 
yet identified the nature of the inhomogeneous phase, we expect that it will be
energetically more favorable, such that
\begin{equation}
2 \pi \rho_s \leq \frac{1}{2} M_\eff \Lambda^2 + 
(G_2 +G_3) \frac{M_\eff^2 \Lambda^2}{8 \pi} 
\ \left\{\begin{array}{l} \mbox{Inhomogeneous phase of} \\ 
\mbox{yet unidentified nature.} \end{array} \right.
\end{equation}
At least at ${\cal O}(G_i)$, the 45 degrees spiral cannot be realized in this
regime, because for $G_i = 0$ it only exists at the isolated point 
$2 \pi \rho_s = \frac{1}{2} M_\eff \Lambda^2$ and first order perturbation 
theory uses just the unperturbed wave function. Definitely, the one-pocket 45 
degrees spiral becomes unstable when $\kappa_1 < 0$, i.e.\ for
\begin{equation}
2 \pi \rho_s \leq \frac{1}{2} M_\eff \Lambda^2
\ \left\{\begin{array}{l} \mbox{Instability of the 45 degrees spiral} 
\\ \mbox{against formation of inhomogeneities.} \end{array} \right.
\end{equation}
It should be pointed out again that these results apply only if the 4-fermion 
contact interactions are weak. If the 4-fermion couplings are strong, the 
approach of filling pockets with weakly interacting holes is not applicable and
the true ground state of the system may be different. Even if the 4-fermion 
couplings are small, the present result does not necessarily reveal the 
complete nature of the true ground state. In particular, the configurations of 
the staggered magnetization were restricted to those that imply a homogeneous 
composite vector field $v_i(x)'$, i.e.\ to spirals or to configurations that 
are homogeneous themselves. For example, the double spiral, which implies an 
inhomogeneous $v_i(x)'$, was not taken into account. Hence, we cannot exclude 
that the phases that we found may still be unstable against developing 
inhomogeneities, at least in certain parameter regions. This shall be explored 
in the future.

\section{Reduction of the Staggered Magnetization \\ upon Doping}

Until now we have considered the Lagrangian or the Hamiltonian of the effective
theory. Now we will see that other quantities can be constructed in a similar 
way. An observable of particular interest is the staggered magnetization which 
is the order parameter for the spontaneous breakdown of the $SU(2)_s$ symmetry.
As the antiferromagnet is doped, the staggered magnetization is reduced until 
the $SU(2)_s$ symmetry is restored. In the actual materials this happens at
relatively small doping, before high-temperature superconductivity sets in.

In the undoped antiferromagnet the local staggered magnetization is given by
$\vec M_s(x) = {\cal M}_s \vec e(x)$. The low-energy parameter ${\cal M}_s$ is 
the length of the staggered magnetization vector. For the Heisenberg model
(or equivalently for the $t$-$J$ model at half-filling) this parameter has been
determined with high precision in quantum Monte Carlo simulations resulting in 
${\cal M}_s = 0.3074(2)/a^2$ \cite{Wie94,Bea96}. In a doped antiferromagnet the
staggered magnetization receives additional contributions from the fermions
such that
\begin{equation}
\vec M_s(x) = \Big[{\cal M}_s - 
m \sum_{\ontopof{f=\alpha,\beta}{\, s = +,-}} \psi^{f\dagger}_s(x) \psi^f_s(x) 
\Big] 
\vec e(x).
\end{equation}
Here $m$ is another low-energy parameter which determines the reduction of the
staggered magnetization upon doping. It should be noted that there are further
contributions to $\vec M_s(x)$ which include derivatives or contain more than 
two fermion fields. All these terms are of higher order and will be neglected 
here. Since both the homogeneous and the spiral phases have a constant fermion 
density we can use
\begin{equation}
\sum_{\ontopof{f=\alpha,\beta}{\, s = +,-}} \langle \Psi^{f\dagger}_s \Psi^f_s 
\rangle = n,
\end{equation}
such that 
\begin{equation}
{\cal M}_s(n) = {\cal M}_s - m n.
\end{equation}
The higher-order terms that we have neglected correct this equation at
${\cal O}(n^2)$. A rough estimate of the critical density at which the 
$SU(2)_s$ symmetry gets restored is $n_c \approx {\cal M}_s/m$. It would be 
interesting to determine the value of $m$ for the $t$-$J$ model, which may  be 
feasible in quantum Monte Carlo calculations.

\section{Conclusions and Outlook}

In this paper we have used a systematic effective field theory for 
antiferromagnetic magnons and holes in order to investigate the propagation of
holes in the background of a staggered magnetization field. We have limited 
ourselves to configurations of the staggered magnetization that are either
homogeneous themselves or that generate a constant composite vector field 
$v_i(x)'$ for the fermions. In both cases, the resulting fermion density is
homogeneous. Our calculations also rely on the assumption that remnant
4-fermion contact interactions between the holes are weak and can be treated
in perturbation theory. This may or may not be the case for a concrete magnetic
material. We like to emphasize again that the effective theory is universal 
and makes model-independent predictions for a large class of magnetic systems. 
Material-specific properties enter the effective theory through the values of 
low-energy parameters such as the spin stiffness $\rho_s$. For large values of 
$\rho_s$ distortions in the staggered magnetization cost a large amount of 
energy and a homogeneous phase is energetically favored. In that case, all four
hole pockets are equally populated with doped holes. For smaller values of 
$\rho_s$, on the other hand, the doped holes can gain energy from a spiral in 
the staggered magnetization. For intermediate values of $\rho_s$ a zero degree 
spiral, in which only two hole pockets are populated, is most stable, while a
45 degrees spiral (with either one or three populated hole pockets) can exist 
only at two specific isolated values of $\rho_s$. It is conceivable that the
45 degrees spirals may be stabilized for larger values of the 4-fermion
couplings $G_i$ (inaccessible to first order perturbation theory). For small 
values of $\rho_s$ a yet unidentified inhomogeneous phase is energetically 
favored. The reduction of the staggered magnetization upon doping is again 
controlled by a low-energy parameter $m$ whose value depends on the material in
question.

The results of our investigation provide a basis for further studies that
naturally suggest themselves. First, it would be interesting to investigate if
there is a stable minimum of the energy for large values of $c_i^3$. This
requires a straightforward but somewhat tedious numerical calculation which we
have not yet performed. Next, instead of using first order perturbation theory,
one may want to include the 4-fermion couplings in a variational calculation.
This would eliminate the assumption that these couplings are small. One should 
also analyze the stability of the
homogeneous and spiral phases against developing inhomogeneities in the 
fermion density. For example, it is interesting to ask if a double spiral is
energetically more favorable than the spiral phases considered here. If this
were the case, it is conceivable that a lightly doped antiferromagnet without
impurities is an insulator. The real materials are indeed insulators, but they
contain impurities on which the doped holes may get localized. 

On the other hand, if the homogeneous phase or the simple spirals considered 
here turn out to be more stable than, for example, a double spiral, the effects
of magnon exchange between doped holes would be interesting to study in detail.
In particular, using the effective theory, we have shown that the one-magnon 
exchange potential between two isolated holes gives rise to binding with 
$d$-wave characteristics \cite{Bru05,Bru06}. In a spiral phase the exchange of 
spiral phonons (i.e.\ helimagnons) may also be an important dynamical mechanism
for the binding of hole pairs. Depending on the size of the pairs and on their 
density, hole  pair formation may lead to Bose-Einstein condensation or to 
magnon-mediated BCS-type superconductivity coexisting with antiferromagnetism. 
While coexistence of superconductivity and antiferromagnetism is not observed 
in the high-$T_c$ cuprates, this may be due to impurities created by doping on 
which holes may get localized, thus preventing superconductivity. A clean 
system like the $t$-$J$ model, on the other hand, may superconduct already at 
small doping within the antiferromagnetic phase, although $T_c$ may then be 
rather small. The effective theory provides us with a tool that allows us to 
address such questions. If there is superconductivity already within the 
antiferromagnetic phase in a clean system, the corresponding mechanism 
responsible for superconductivity may persist in the high-$T_c$ cuprates, 
despite the fact that in the real materials at small doping superconductivity 
may be prevented by the localization of holes on impurities. Of course, once 
antiferromagnetism is destroyed, magnons and spiral phonons no longer exist as 
massless excitations. However, antiferromagnetic correlations, although only of
finite range, still exist in high-$T_c$ superconductors and may still play an 
important role as relevant low-energy degrees of freedom. This is similar to 
nuclear physics where the pion is not exactly massless (in that case due to 
explicit chiral symmetry breaking) but is certainly relevant at low energies.
Still, whether magnon-mediated binding between holes may be responsible for 
high-temperature superconductivity remains a difficult question that may or may
not be within the applicability range of the effective theory. For lightly 
doped cuprates the low-energy effective field theory is applicable. It yields 
reliable and interesting results and should be pursued further.

\section*{Acknowledgements}

C.\ P.\ H. would like to thank the members of the Institute for Theoretical 
Physics at Bern University for their hospitality during a visit at which this 
project was initiated. U.-J.\ W. likes to thank P.\ A.\ Lee for interesting 
discussions and the members of the Center for Theoretical Physics at MIT, where
this work was completed, for their hospitality. This work was supported in part
by funds provided by the Schweizerischer Nationalfonds. The work of C.\ P.\ H.\
is supported by CONACYT grant No. 50744-F, by Fondo Ram\'on Alvarez-Buylla de 
Aldana grant No.\ 349/05, and by grant PIFI 3.2.

\begin{appendix}

\section{Most General Configuration with a \\ 
Homogeneous Composite Vector Field}

In this appendix we show that the most general configuration with constant
$v_i(x)'$ corresponds to a spiral in the staggered magnetization. According to
eq.(\ref{const}) we need to consider $\theta(x)$, $\varphi(x)$, and $\alpha(x)$
such that
\begin{eqnarray}
{v^3_i}(x)'&=&v^3_i(x) - \p_i \alpha(x) = 
\sin^2\frac{\theta(x)}{2} \p_i \varphi(x) - \p_i \alpha(x) = c^3_i,
\nonumber \\
{v^\pm_i}(x)'&=&v^\pm_i(x) \exp(\pm 2 i \alpha(x)) \nonumber \\
&=&\frac{1}{2} (\sin\theta(x) \p_i \varphi(x) \pm i \p_i \theta(x)) 
\exp(\mp i(\varphi(x) - 2 \alpha(x))) = c^\pm_i,
\end{eqnarray}
with $c_i^\pm$ and $c_i^3$ being constant. Introducing 
$\chi(x) = \varphi(x) - 2 \alpha(x)$ one obtains
\begin{eqnarray}
\label{dalpha}
&&\cos\theta(x) \p_i \alpha(x) = \sin^2\frac{\theta(x)}{2} \p_i \chi(x) - 
c^3_i, 
\nonumber \\
&&\tan\theta(x) (\p_i \chi(x) - 2 c^3_i) \pm i \p_i \theta(x) = 2 c^\pm_i 
\exp(\pm i \chi(x)) = 2 c_i \exp(\pm i (\chi(x) + \omega_i)). \nonumber \\ \,
\end{eqnarray}
Here we have put
\begin{equation}
c^\pm_i = c_i \exp(\pm i \omega_i), \quad  \mbox{with} \ c_i \in \R, \quad
\omega_i \in [- \frac{\pi}{2},\frac{\pi}{2}].
\end{equation}
By an appropriate constant gauge transformation, one can always put 
$\omega_1 = 0$. From eq.(\ref{dalpha}) one infers
\begin{eqnarray}
&&\tan\theta(x) (\p_i \chi(x) - 2 c^3_i) = 2 c_i \cos(\chi(x) + \omega_i),
\nonumber \\
&&\p_i \theta(x) = 2 c_i \sin(\chi(x) + \omega_i).
\end{eqnarray}
Demanding $\p_1 \p_2 \theta(x) = \p_2 \p_1 \theta(x)$ leads to the constraints
\begin{equation}
\label{a}
\frac{c_i}{c^3_i} = a, \quad \omega_1 = \omega_2 = 0,
\end{equation}
which imply
\begin{eqnarray}
&&\p_i \chi(x) = 2 c^3_i[1 + a \cot\theta(x) \cos\chi(x)],
\nonumber \\
&&\p_i \theta(x) = 2 c^3_i a \sin\chi(x).
\end{eqnarray}
These equations can be satisfied only if $\chi$ and $\theta$ are 
functions of $z = c^3_i x_i$, i.e.\ if they are plane waves with
\begin{eqnarray}
\label{dz}
&&\p_z \chi(z) = 2 [1 + a \cot\theta(z) \cos\chi(z)],
\nonumber \\
&&\p_z \theta(z) = 2 a \sin\chi(z).
\end{eqnarray}
Since both $\chi$ and $\theta$ depend on $z$ only, we may also consider $\chi$
as a function of $\theta$ which then leads to
\begin{equation}
\p_\theta \chi(\theta) = \frac{1}{a \sin\chi(\theta)} + 
\cot\theta \cot\chi(\theta).
\end{equation}
This can be cast into the form
\begin{equation}
- \p_\theta \cos\chi(\theta) = 
\frac{1}{a} + \cot\theta \cos\chi(\theta),
\end{equation}
and is solved by
\begin{equation}
\label{coschi}
\cos\chi(\theta) = \frac{\cos\theta - \lambda}{a \sin\theta},
\end{equation}
where $\lambda$ is an integration constant. Inserting this result in 
eq.(\ref{dz}) one obtains
\begin{eqnarray}
\label{diff}
&&- \p_z \cos\theta(z) = \pm
2 \sqrt{a^2(1 - \cos^2\theta(z)) - (\cos\theta(z) - \lambda)^2}, \nonumber \\
&&\p_z \varphi(z) = \frac{2 (\cos\theta(z) - \lambda)}{\sin^2\theta(z)}.
\end{eqnarray}
The equation for $\p_z \varphi(z)$ results by combining eq.(\ref{dalpha})
with eq.(\ref{dz}) and eq.(\ref{coschi}). The above equation for 
$\cos\theta(z)$ can be integrated in a straightforward manner and one obtains
\begin{eqnarray}
\cos\theta(z)&=&\frac{1}{1 + a^2}\left[\lambda + a \sqrt{1 + a^2 - \lambda^2}
\cos\left(2 \sqrt{1 + a^2} (z - z_0)\right)\right] \nonumber \\
&=&\frac{1}{\sqrt{1 + a^2}}\left[\cos\eta + a \sin\eta 
\cos\left(2 \sqrt{1 + a^2} (z - z_0)\right)\right].
\end{eqnarray}
Here we have expressed the integration constant as
\begin{equation}
\lambda = \sqrt{1 + a^2} \cos\eta.
\end{equation}
It will turn out that $\eta$ is the angle between the direction $\vec j$
perpendicular to the spiral plane and the 3-axis. Furthermore, one obtains
\begin{eqnarray}
&&\sin\theta(z) \cos(\varphi(z) - \varphi(z_0)) =
\frac{1}{\sqrt{1 + a^2}}\left[\sin\eta - a \cos\eta 
\cos\left(2 \sqrt{1 + a^2} (z - z_0)\right)\right], \nonumber \\
&&\sin\theta(z) \sin(\varphi(z) - \varphi(z_0)) =
\frac{a}{\sqrt{1 + a^2}} \sin\left(2 \sqrt{1 + a^2} (z - z_0)\right),
\end{eqnarray}
and thus
\begin{equation}
\varphi(z) - \varphi(z_0) = \mbox{atan}
\left(\frac{a \sin\left(2 \sqrt{1 + a^2} (z - z_0)\right)}
{\sin\eta - a \cos\eta \cos\left(2 \sqrt{1 + a^2} (z - z_0)\right)}\right).
\end{equation}
Differentiating this equation with respect to $z$, it is straightforward to
show that eq.(\ref{diff}) is indeed satisfied.

We now form the scalar product of the unit-vector
\begin{equation}
\vec e(z) = (\sin\theta(z) \cos\varphi(z),\sin\theta(z) \sin\varphi(z),
\cos\theta(z)),
\end{equation}
describing the staggered magnetization, with the unit-vector
\begin{equation}
\vec j = (\sin\eta \cos\varphi(z_0),\sin\eta \sin\varphi(z_0),\cos\eta),
\end{equation}
which yields
\begin{eqnarray}
\vec e(z) \cdot \vec j&=&\sin\theta(z) \sin\eta 
(\cos\varphi(z) \cos\varphi(z_0) + \sin\varphi(z) \sin\varphi(z_0)) +
\cos\theta(z) \cos\eta \nonumber \\
&=&\sin\theta(z) \sin\eta \cos(\varphi(z) - \varphi(z_0)) +
\cos\theta(z) \cos\eta = \frac{1}{\sqrt{1 + a^2}},
\end{eqnarray}
i.e.\ a constant ($z$-independent) scalar product. This finally proves that 
$\vec e(z)$ indeed describes a spiral in a plane perpendicular to $\vec j$.
Replacing $z = c_i^3 x_i$ and using eq.(\ref{a}), the wave number of the spiral
(i.e.\ the spiral pitch) can be identified as
\begin{equation}
\label{pitch}
k = 2 \sqrt{1 + a^2} \sqrt{c_i^3 c_i^3} = 2 \sqrt{c_i^3 c_i^3 + c_i c_i}.
\end{equation}

The particular spiral configuration considered in eq.(\ref{special}) has
$\eta = 0$ such that
\begin{eqnarray}
&&\cos\theta(z) = \frac{1}{\sqrt{1 + a^2}} = \cos\theta_0, \nonumber \\
&&\sin\theta(z) \sin(\varphi(z) - \varphi(z_0)) =
\frac{a}{\sqrt{1 + a^2}} \sin\left(2 \sqrt{1 + a^2} (z - z_0)\right) 
\Rightarrow \ \nonumber \\
&&\varphi(z) - \varphi(z_0) = - 2 \sqrt{1 + a^2} \ c_i^3 x_i = k_i x_i.
\end{eqnarray}
This is indeed consistent with eq.(\ref{ci3}) which implies
\begin{equation}
a = - \tan\theta_0, \quad k_i = - \frac{2 c_i^3}{\cos\theta_0}.
\end{equation} 

\end{appendix}

\end{document}